\documentclass[fleqn,usenatbib]{mnras}

\usepackage{newtxtext,newtxmath}

\usepackage[T1]{fontenc}

\DeclareRobustCommand{\VAN}[3]{#2}
\let\VANthebibliography\thebibliography
\def\thebibliography{\DeclareRobustCommand{\VAN}[3]{##3}\VANthebibliography}

\usepackage{graphicx}	
\usepackage{amsmath}	

\usepackage{comment}

\title[Neutron stars in supernova remnants]{Initial periods and magnetic fields of neutron stars   }

\author[Andrei P. Igoshev et al.]{
Andrei P. Igoshev,$^{1}$\thanks{E-mail: ignotur@gmail.com a.igoshev@leeds.ac.uk}
Anastasia Frantsuzova,$^{2}$
Konstantinos N. Gourgouliatos$^{3}$,
Savina Tsichli$^{3}$, 
\newauthor Lydia Konstantinou$^{3}$
and Sergei B. Popov$^{4}$
\\
$^{1}$Department of Applied Mathematics, University of Leeds, LS2 9JT Leeds, UK\\
$^{2}$School of Mathematics, University of Leeds, LS2 9JT Leeds, UK\\
$^{3}$Department of Physics, University of Patras, 26504 Patras, Greece\\
$^{4}$Sternberg Astronomical Institute, Lomonosov Moscow State University, Moscow, 119234 Russia\\
}

\date{Accepted XXX. Received YYY; in original form ZZZ}

\pubyear{2015}

\begin{document}
\label{firstpage}
\pagerange{\pageref{firstpage}--\pageref{lastpage}}
\maketitle

\begin{abstract}
Initial distributions of pulsar periods and magnetic fields are essential components of multiple modern astrophysical models. Not enough work has been
done to properly constrain these distributions using direct measurements. Here we aim to fill this gap by rigorously analysing properties of young neutron stars associated to supernova remnants. In order to perform this task, we compile a catalogue of 56 neutron stars uniquely paired to supernova remnants with known age estimate. Further, we analyse this catalogue using multiple statistical techniques. We found that distribution of magnetic fields and periods for radio pulsars are both well described using the log-normal distribution. The mean magnetic field is $\log_{10} [B/\mathrm{G}] = 12.44$ and standard deviation is $\sigma_B = 0.44$. Magnetars and central compact objects do not follow the same distribution. 
The mean initial period is $\log_{10} P_0 [P / \mathrm{s}] = -1.04_{-0.2}^{+0.15}$ and standard deviation is $\sigma_p = 0.53_{-0.08}^{+0.12}$. We show that the normal distribution does not describe the initial periods of neutron stars sufficiently well. Parameters of the initial period distribution are not sensitive to the exact value of the braking index.  
\end{abstract}

\begin{keywords}
stars: neutron -- methods: statistical -- pulsars: general
\end{keywords}



\section{Introduction}

Knowledge of initial properties of neutron stars (NSs) is essential for the understanding of high-energy astrophysical phenomena, in particular gamma-ray bursts (GRBs; for a review see, e.g. \citealt{ZhangReview2004, Schady2017}) and fast radio bursts (FRBs; for a review see \citealt{PetroffHessels2019, 2022arXiv220314198X}). Thus, it has been suggested that newly born strongly magnetised NSs (magnetars $B\sim 10^{14}$~--~$10^{15}$~G) with fast rotation (a few msec periods) could be responsible for GRBs afterglow \citep{Usov1992Natur,DuncanThompson1992ApJ,DallOsso2011A,Rowlinson2010}. Magnetars are identified as central engine for at least one FRB  thanks to simultaneous detection of a radio \citep{2020Natur.587...54C, 2020Natur.587...59B} and high energy \citep{2020ApJ...898L..29M, 2021NatAs...5..378L, 2021NatAs...5..372R, 2021NatAs...5..401T}  bursts from a Galactic source SGR 1935+2154.
Additionally, the distribution of initial periods is important to constrain pulsar ages e.g. to study energy sources of older radio pulsars (see \citealt{Abramkin2021arXiv}).

The distribution of initial NS magnetic field is crucial for the study of the formation of strong magnetic fields during NS collapse and convection at the proto-NS stage \citep{Makarenko2021MNRAS}. Magnetic fields of NSs significantly determine their observational appearances \citep{IgoshevReview2021}, and thus are important ingredients of pulsar population syntheses, see e.g. \cite{FaucherGiguere2006ApJ,PopovPons2010MNRAS,GullonPons2014MNRAS}.

The most likely channel for the formation of a NS is that of a core-collapse supernova explosion \citep{2018ASSL..457....1C} of stars with masses of $M>8 M_{\odot}$, where possible outcomes are either a black hole for the more massive progenitors or an NS for the less massive ones, but with a dependence on other properties of the progenitors such as metallicity and rotation \citep{2003ApJ...591..288H}. The ages of  those supernova remnants can be estimated through their expansion rate and sizes. Some NSs could be formed as a result of electron-capture supernova explosion \citep{Miyaji80,Nomoto1984,NomotoKondo91,WoosleyHeger15,Jones16}. At the moment, it is unknown whether these NSs have significantly different initial periods and magnetic fields. These NSs might form less noticeable (faster disappearing) SNR. This possibility was discussed by \cite{2021MNRAS.508..279C}. We review the significance of this prospect for our own research in Section~\ref{s:alikepsr}.  

The moment that an NS is born is obscured by dense stellar material, thus immediate properties of newly born NS are impossible to measure. Future gravitational-wave observatories will probably constrain some of the proto-NS properties \citep{Radice2019ApJ}.
At the moment it is only possible to constrain the initial properties of NSs by analysing a sample of the youngest NSs associated to supernova remnants (SNRs). These remnants are relatively easy to find in radio and/or X-ray observations and they stay bright and structured up to $\approx 30$~kyr age \cite{2008ARA&A..46...89R}. Only some SNRs host a known radio pulsar. There are a few explanations for this lack of NSs inside SNRs: (1) the sensitivity of modern radio survey is thought to be insufficient to detect some radio pulsars \citep{Sett2021}; (2) due to beaming we could miss radio emission from significant fraction of radio pulsars; (3) SNR could have been formed as a result of SN Ia which does not produce NS; or (4) SNR could be associated to birth of a black hole.

A number of studies have already concentrated on young NSs associated to SNRs. E.g., \cite{PopovTurolla2012} analysed information about $\sim30$ NSs associated to SNR and found that the distribution of NS periods can be roughly described using the normal distribution with mean $\mu\sim0.1$~s and $\sigma \sim 0.1$. This distribution has certain non-physical properties. For example, its tail can theoretically extend to negative periods (a fraction of approx. 15~per~cent). In practical applications these negative periods are removed, but initial period distribution should intrinsically be described using functions which are defined only at the positive values.

In this article we match the SNR catalogue \citep{Ferrand2012} with ATNF pulsar catalogue\footnote{\url{http://www.atnf.csiro.au/research/pulsar/psrcat}} \citep{ATNF} and identify 68 NSs with possible association to SNRs. Many of these pairs have been identified already in the literature. We further investigate the distributions of magnetic fields and periods. 

Throughout most of the paper we assume that the sample of NSs associated to supernova remnant is representative of the population of young NSs. It means that parameters which we estimate for this sample are representative of parameters for young neutron star population. It might not be the case due to some observational selection, which we discuss in Section~\ref{s:selection}.

The article is structured as the following. In Section~\ref{s:data} we describe how we identify NSs associated with SNRs, in Section~\ref{s:analysis} we analyse the distribution of magnetic fields and computed initial periods. 
In Section~\ref{s:linear_model_B_evol} we study a linear model for magnetic field evolution. We discuss our results in Section \ref{DISCUSSION} and we conclude in Section \ref{CONCLUSION}.

\section{Data}
\label{s:data}

\subsection{Data acquisition}

\begin{figure}
	\includegraphics[width=\columnwidth]{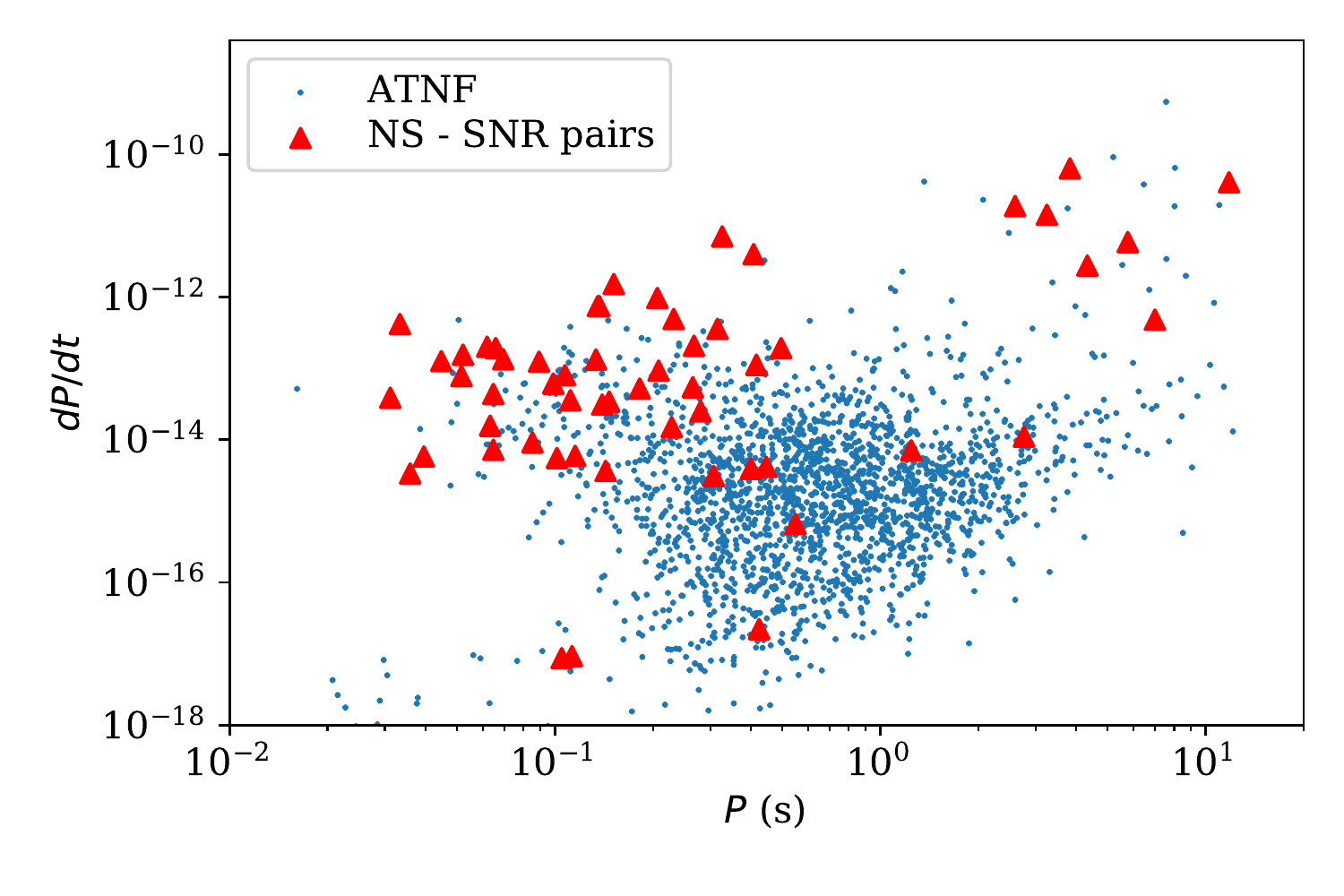}
    \caption{Distribution of neutron stars associated to supernova remnants (red triangles) and radio pulsars (blue dots) over period and period derivatives. Periods and period derivatives of radio pulsars are obtained from the ATNF catalogue v.1.66. Period and period derivatives for NS associated to SNR are collected from Table~\protect\ref{t:catalogue}.  }
    \label{fig:pdotp}
\end{figure}

We have used data presented in the SNRcat\footnote{\url{ http://snrcat.physics.umanitoba.ca}} \citep{Ferrand2012}. The catalogue contains 383 records of SNRs. Out of these, 295 are also included in the Green catalogue\footnote{\url{https://www.mrao.cam.ac.uk/surveys/snrs/snrs.info.html}} \citep{2019JApA...40...36G}, while the remaining 88 sources include also candidate sources. There are 133 associations with neutron stars, either confirmed or candidate, out of which 110 are pulsars. We have selected the set of sources under the following criteria. Firstly, there exist a measurement of period and period derivative of the pulsar, so that we can estimate the intensity of the magnetic dipole field and the characteristic age of the hosted pulsar. Secondly, there is an estimate of the SNR age due to expansion. We have found 68 sources that fulfil these conditions, which are presented in the Appendix \ref{appendix1}. 

In this table we also summarise the fundamental properties of radio pulsars, such as period $P$ and period derivative $\dot P$. 
%
%
For each NS associated to SNR we add a label describing NS type. We use the following designations: isolated radio pulsar (PSR), anomalous X-ray pulsar (AXP), soft gamma repeater (SGR), central compact object (CCO) or high-B radio pulsar (HBRP). We choose the AXP and SGR labels using the information from the McGill Online Magnetar Catalog\footnote{\url{http://www.physics.mcgill.ca/~pulsar/magnetar/main.html}} \citep{Olausen2014ApJS}. If a particular magnetar is observed as a radio pulsar, we mark it as HBRP, as in the case of PSR J1622-4950.

\subsection{Unique NS -- SNR pairing}

The majority of these sources are located close to the Galactic plane, thus there is some overlap between the sources simply by chance. 
Because of this, in our initial catalogue of NSs associated to SNRs (Table~\ref{t:catalogue}) we have two types of association difficulties: (1) in some cases multiple NSs are located close to a single SNR and (2) in some cases multiple SNRs are located close to a single NS. A remarkable example of first issue is SNR G011.2-00.3, which is located close to three young NSs J1811-1925, J1809-1917, and J1809-1943. To give an example of the second problem, we point toward PSR J1640-4631 which is close to SNR G338.3-00.0 and SNR G338.5+00.1. 

To address the first issue we select the NS, among all candidates, whose spin-down age is closest to the SNR age. Thus, in our analysis we include only SNR G011.2-00.3 paired with J1811-1925. In order to solve the second issue, we choose SNR with age estimate closest to the NS spin-down age. Thus we include only SNR G338.3-00.0 paired with PSR J1640-4631. The information about included and excluded pairs is provided in Table~\ref{t:catalogue} in column 'Included'.
We make a single exception from this rule. In the case of SNR G033.6+00.1 we assume that it is paired with CCO J1852+0040 rather than with the SGR 3XMM J185246.6+003317 as CCO-like sources are never detected outside remnants, while SGR are often seen without a SNR. 
After this cleaning procedure, our final catalogue includes 56 NSs uniquely paired with 56 SNRs. 

We show location of NSs at the period -- period derivative diagram in Figure~\ref{fig:pdotp}. Radio pulsars associated to supernova remnants are numerous in the upper left corner of the diagram and much less frequent at periods around 1~sec. It reassures the general view that pulsars in their rotational evolution move from upper left corner to lower middle part of period-period derivative diagram. The NSs in the upper right corner are magnetars.

\section{Analysis}
\label{s:analysis}

\subsection{Magnetic field distribution of neutron stars}

For our analysis we divide all pairs NS-SNR into two groups: (1) all NS and (2) only radio pulsars excluding HBPSR. We consider an NS to be a radio pulsar if it shows radio emission typical for normal radio pulsars and never demonstrated any activity associated to magnetars. 
We further perform exactly the same analysis for each of these two groups. 

The spin evolution of an isolated NS under influence of electromagnetic torques 
could be described as the following:
\begin{equation}
\dot E_\mathrm{rot}=I\Omega \dot \Omega = - K I \Omega^{n+1}  ,
\label{eq:mag-dip}
\end{equation}
where $\Omega=2 \mathrm{\pi} /P$ is the angular frequency, $I$ is moment of inertia, $K$ is a constant, and $n$ is the braking index (see, e.g. \citealt{Michel1970ApL}). In modern numerical simulations \citep{Philippov2014MNRAS} it is found that:
\begin{equation}
P\frac{dP}{dt}= (\kappa_0 + \kappa_1 \sin^2 \chi) B_p^2 \beta ,    
\label{pdotp_beta}
\end{equation}
where $\chi$ is the obliquity angle between the rotational axis of the radio pulsar and orientation of its dipolar magnetic field with polar strength $B_p$. Values of coefficients $\kappa_0$ and $\kappa_1$ are estimated in numerical MHD simulations by \cite{Philippov2014MNRAS} as $\kappa_0\approx 1$ and $\kappa_1\approx 1.2$.
The constant $\beta$ is computed as:
\begin{equation}
\beta = \frac{\pi^2 R^6}{c^3 I}  .  
\end{equation}
Here $c$ is the speed of light, $R$ is the NS radius, and $I$ is the inertia moment. 

Oftentimes, poloidal, dipolar magnetic fields are estimated \citep{Lorimer2012book} using spin period $P$ and period derivative $\dot P$ as:
\begin{equation}
B_\mathrm{p} = 3.2\times 10^{19} \sqrt{P\dot P} \; \mathrm{G}.
\label{eq:bp}
\end{equation}
This equation is derived under assumptions of magneto-dipole spin-down with angle $\chi=\mathrm{const}=90^\circ$ between spin and magnetic axes. 
Below we use this simplified approach to model spin-down evolution, keeping $B_\mathrm{p}$ and $\chi$ constant, except Section~\ref{s:linear_model_B_evol}. Note that other spin evolution laws have been proposed, see e.g. \cite{2020MNRAS.494.3899N} and references therein. 

Both period and period derivative are measured with large precision: period is typically known with relative error of $10^{-14}$ and period derivative with relative error of $10^{-6}$, which translates to negligible uncertainty in $B_p$ when other parameters fixed.
Therefore, it is possible to use the sample mean of $B_p$ as an estimate of the population mean and sample standard deviation to estimate the standard deviation for the population. We discuss observational selection in Section~\ref{s:selection}.

In a similar manner we define characteristic age of a pulsar:
\begin{equation}
\tau = \frac{P}{2\dot{P}}\,,
\label{eq:tau_ch_age}
\end{equation}
assuming that it spins down due to dipole magnetic field. We note that the actual values of the physical quantities do not affect the characteristic age, provided they remain constant. 

\begin{figure*}
    \begin{minipage}{0.49\linewidth}
	\includegraphics[width=\columnwidth]{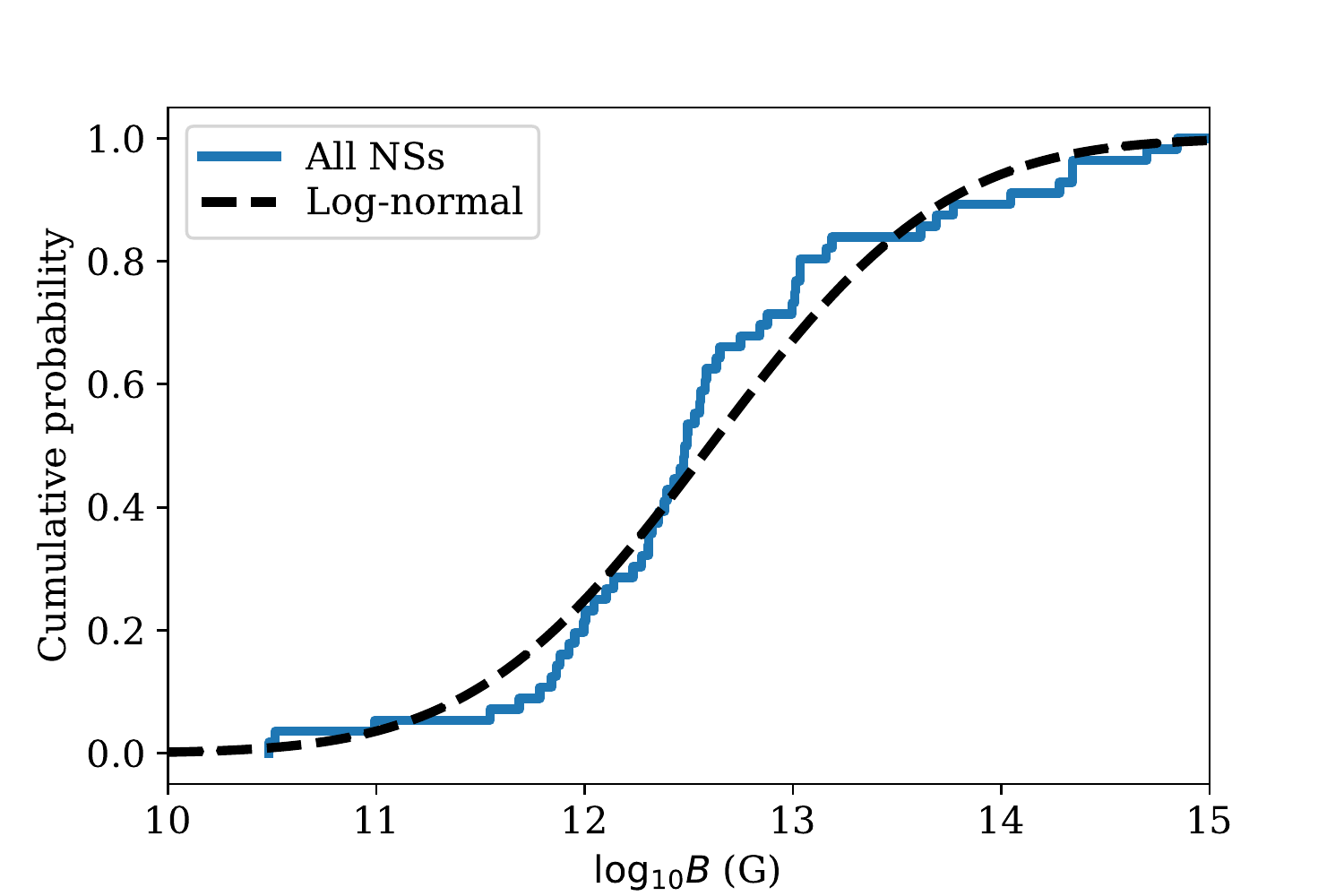}
	\end{minipage}
	\begin{minipage}{0.49\linewidth}
	\includegraphics[width=\columnwidth]{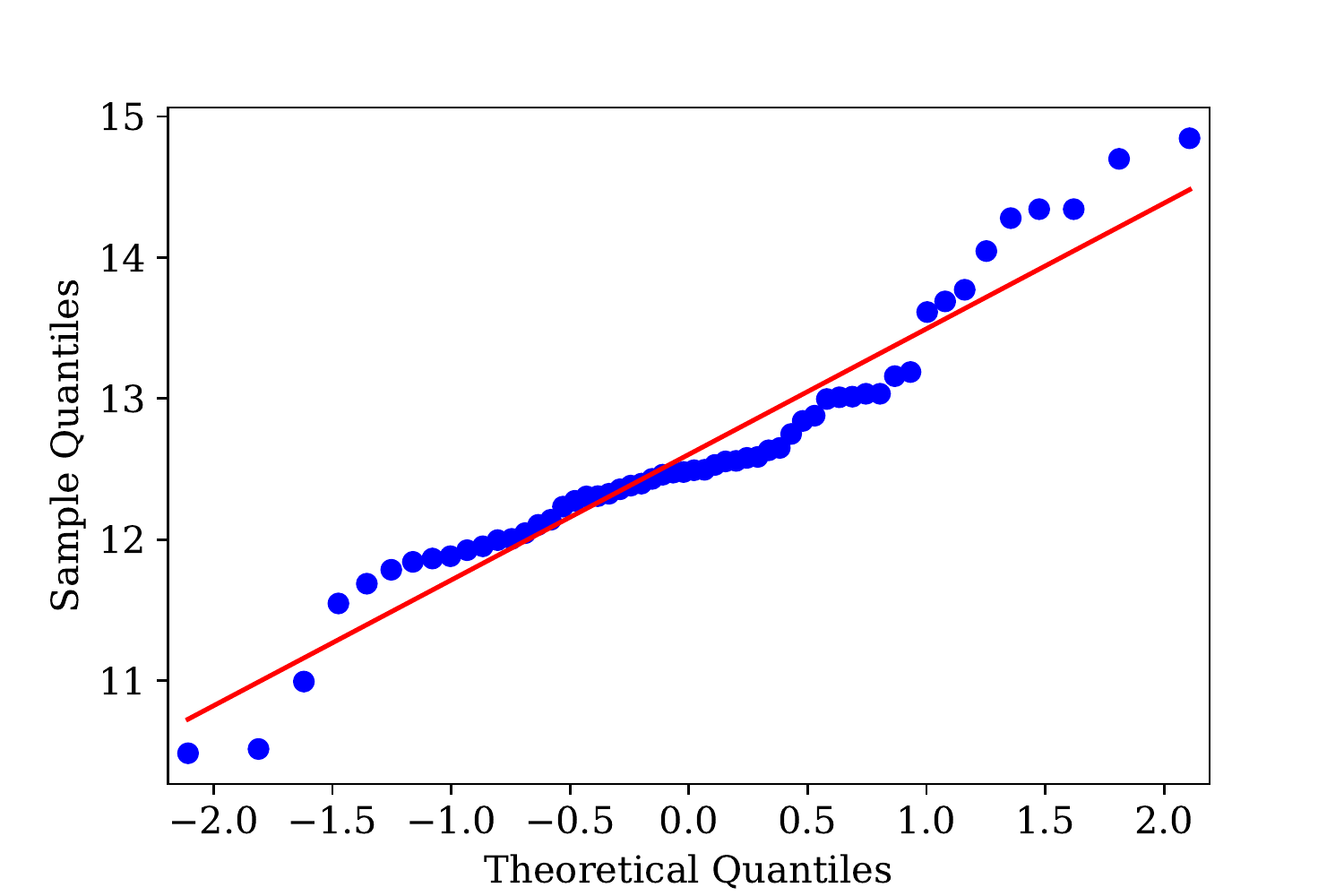}
	\end{minipage}
	\begin{minipage}{0.49\linewidth}
	\includegraphics[width=\columnwidth]{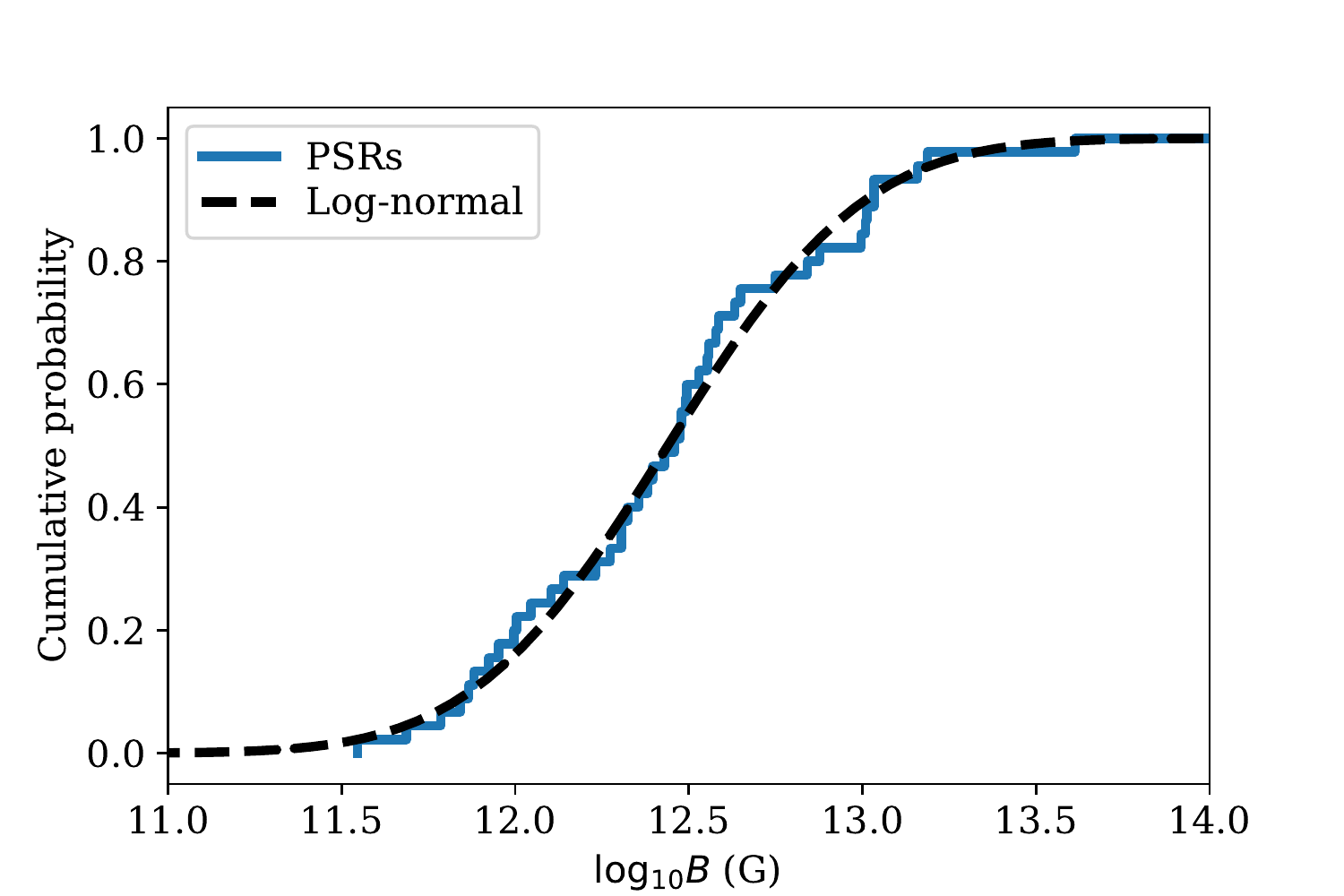}
	\end{minipage}
	\begin{minipage}{0.49\linewidth}
	\includegraphics[width=\columnwidth]{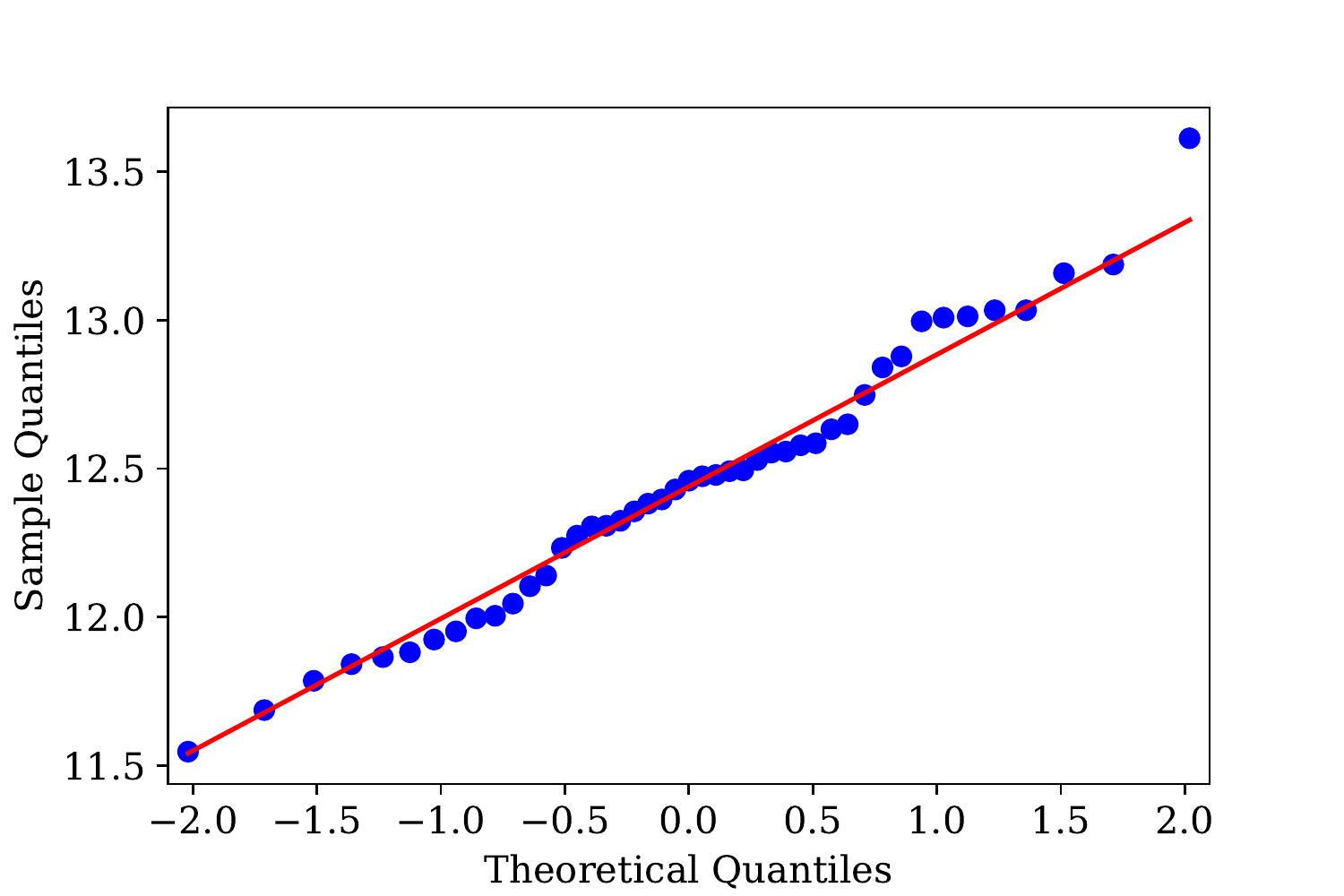}
	\end{minipage}
    \caption{Left panel: cumulative distribution of NSs (upper panel) and radio pulsars (lower panel) within SNRs with logarithmic transformation of their magnetic field estimated via timing. Right panel: Q-Q plot for log-normal distribution for magnetic fields of NSs (upper panel) and normal radio pulsars (lower panel).
    Dashed black line is a log-normal distribution with $\mu_B = 12.59$ and $\sigma = 0.86$ (all NSs) and $\mu_B = 12.42$ and $\sigma = 0.46$ (PRSs). Red is reference line. }
    \label{fig:logB_distr}
\end{figure*}

We plot the cumulative distribution of base-10 logarithms for NS magnetic fields in Figure~\ref{fig:logB_distr}. The top plot shows the distribution for all NSs including magnetars and CCOs. The bottom plot includes only radio pulsars. It can be noticed by comparing these plots that the magnetic fields of radio pulsars range from  $\approx 3\times 10^{11}$~G to $4\times 10^{13}$~G. Strongly magnetised NSs are magnetars and weakly magnetised NSs are CCOs.

We summarise the mean value for base-10 logarithm of magnetic field and respective standard deviation for both groups in Table~\ref{tab:logB_res}. While mean values are approximately similar, the standard deviation for all NSs is two times larger than the one for PSRs. These large standard deviations are necessary to explain the broader range of magnetic fields seen in magnetars and CCOs. 

In order to check if the log-normal distribution (frequently used in population synthesis see e.g. \citealt{FaucherGiguere2006ApJ}\footnote{\cite{PopovPons2010MNRAS} used Gaussian-in-log distribution which is quite similar to the log-normal distribution}) provides a good model for the initial distribution of NS magnetic fields, we use the Shapiro test. Shapiro test (also known as Shapiro–Wilk test) is a frequentist test used to check if a data set could be modelled by normal distribution.
We summarise the $p$-values  of this test in the last column of Table~\ref{tab:logB_res}. It is easy to see that the initial magnetic fields of radio pulsars are well described by the log-normal distribution. As for the all NSs, suitability of the log-normal distribution is rejected at 3\%  significance level. The reason for this could be seen in Figure~\ref{fig:logB_distr}. While the analytical cumulative distribution function follows closely the histogram of magnetic fields for radio pulsars, there are large deviations between analytical cumulative distribution and histogram for the entire set of NSs. These deviations are most noticeable around magnetic fields in range $2\times 10^{12}$~--~$2\times 10^{13}$~G.

\begin{table}
    \centering
    \caption{Results of analysis of current magnetic fields (top part) and initial periods (lower part) for a sample which includes all NSs associated with SNRs and for radio pulsars only. In all cases $p$-value corresponds to log-normal distribution.}
    \label{tab:logB_res}
    \begin{tabular}{lcccc}
    \hline
    \multicolumn{5}{c}{Magnetic fields} \\
    \hline
    Sample  & N  &  $\log_{10} B$ & $\sigma_B$ &  $p$-value \\
    \hline
    All NSs & 56 & 12.60 & 0.89 & 0.014 \\
    PSRs    & 45 & 12.44 & 0.44 & 0.727 \\
    \hline
    \multicolumn{5}{c}{Initial spin periods} \\
    \hline
    Sample & N & $\log_{10} P$ & $\sigma_p$ & $p$-value \\
    \hline
    All NSs       & 56 & -1.25        & 0.99     & 0.0014 \\
    PSRs          & 45 & -1.34        & 0.81     & 0.0019 \\
    Selected PSRs & 35 & -1.04        & 0.42     & 0.4529 \\
    \hline
    \end{tabular}
\end{table}

Another way to compare the log-normal distribution with observations is to examine the quantile-quantile (Q-Q) plot, see Figure~\ref{fig:logB_distr} right panels. The logarithm of magnetic fields for radio pulsars follows nicely the straight line with an exception of the last measurement. It is expected if the initial distribution for the sample corresponds to the log-normal distribution. On contrary, logarithm of magnetic fields for all NSs deviates from the straight line consistently around the lowest values $B\sim 10^{11}$~G and around the strongest fields $B\sim 10^{14}$~G.

\subsection{Initial period distribution of neutron stars}

In the previous section we analyse NS magnetic fields assuming that they do not show much change on timescales shorter than few $10^5$~years. Although it is generally justified because we observe much older isolated radio pulsars with fields $\approx 10^{12}$~G, we still test this assumption in Section~\ref{s:linear_model_B_evol}. We cannot develop a more precise analysis at the moment, because the magnetic field evolution is not completely understood for young NS. On the contrary, the spin period evolution is much better understood, and there is a general agreement on how spin period changes with time, see eq. (\ref{pdotp_beta}) and \cite{Lorimer2012book}. That is why in our analysis of NS periods we try to restore the initial spin periods.

There are two main uncertainties in the analysis of initial period distribution: (1) the ages of SNRs are not known exactly and often only a range of possible ages is available (see e.g. \citealt{Suzuki2021ApJ}), (2) it is unclear if all pulsars slow down with braking index $n=3$. These are uncertainties of very different nature. In the first case, we know the range of possible SNR ages, thus this uncertainty could be quantified. In the second case, exact $n$ value and its uncertainty is unknown for majority of sources in our catalogue\footnote{This second case encompasses epistemic (systematic) lack of knowledge in comparison to aleatory (statistical) randomness present in the first case.}. Moreover, it is known that the braking index could change during glitches and braking indexes measured on short time scales (months) might significantly differ from effective braking indexes measured on years timescale \citep{Espinoza2017MNRAS}. We compute the initial periods for different constant braking indexes in Section~\ref{s:non3n} 

Because there is an quantifiable uncertainty related to SNR age, we perform our analysis twice using two different techniques. In the first simplified attempt, we assume that true SNR age is the middle of SNR age interval. For example, for J0002+6216 the SNR G116.9+00.2 age interval ranges from 7.5 to 18.1 Kyr. We use the middle of this interval 12.8 Kyr to compute the initial spin period, if it is possible. Further, we employ the classical frequentist toolkit similar to that we used in the previous section to test hypotheses about different shapes of initial spin period distributions for NSs.     

Our second complete analysis is much more mathematically involved and can be skipped by readers unfamiliar with the likelihood technique. These readers can proceed with Section~\ref{s:linear_model_B_evol}. We expect the result of this more complicated analysis to be much more precise, but these results should be in agreement with results of the simpler frequentist analysis. Our main motivation to perform this type of analysis is that some NSs have imaginary (i.e. square root of negative value) values of initial period, if we compute these spin periods using the middle of the SNR age range. Thus we exclude these objects from the simplified analysis. Our likelihood analysis does not have this deficiency. Thus, the sample size is increased.

\subsubsection{Simplified analysis}

We compute the initial periods of radio pulsars using the following equation:
\begin{equation}
P_0 = \sqrt{P^2 - 2 P \dot P t_\mathrm{SNR}}
\label{eq:p0}
\end{equation}
This equation can be derived if we begin with eq. (\ref{eq:mag-dip}).
This procedure means that we assume the braking index to be $n=3$, thus the obliquity angle and magnetic field do not change with time.

Sometimes the expression under the square root of eq. (\ref{eq:p0}) is negative. It could mean different things, e.g. (1) that pulsar was slowing down less efficiently in the past, (2) its SNR age is overestimated, or (3) the simplified model of spin-down is not applicable. Initially, we replace these imaginary periods ($P_0^2 < 0$) with a short period of 0.002~s which is achievable by NSs e.g. the shortest period found in observations (PSR J1748-2446ad with $P = 1.4$~ms; \citealt{Hessels2006Sci}) and a few times slower than typical theoretical estimates for the shortest period of rotation (0.288 ms; \citealt{Haensel1999}; although the crust breaking might limit up to $\approx 1$~ms \citealt{Fattoyev2018arXiv}). 

\begin{figure}
	\includegraphics[width=\columnwidth]{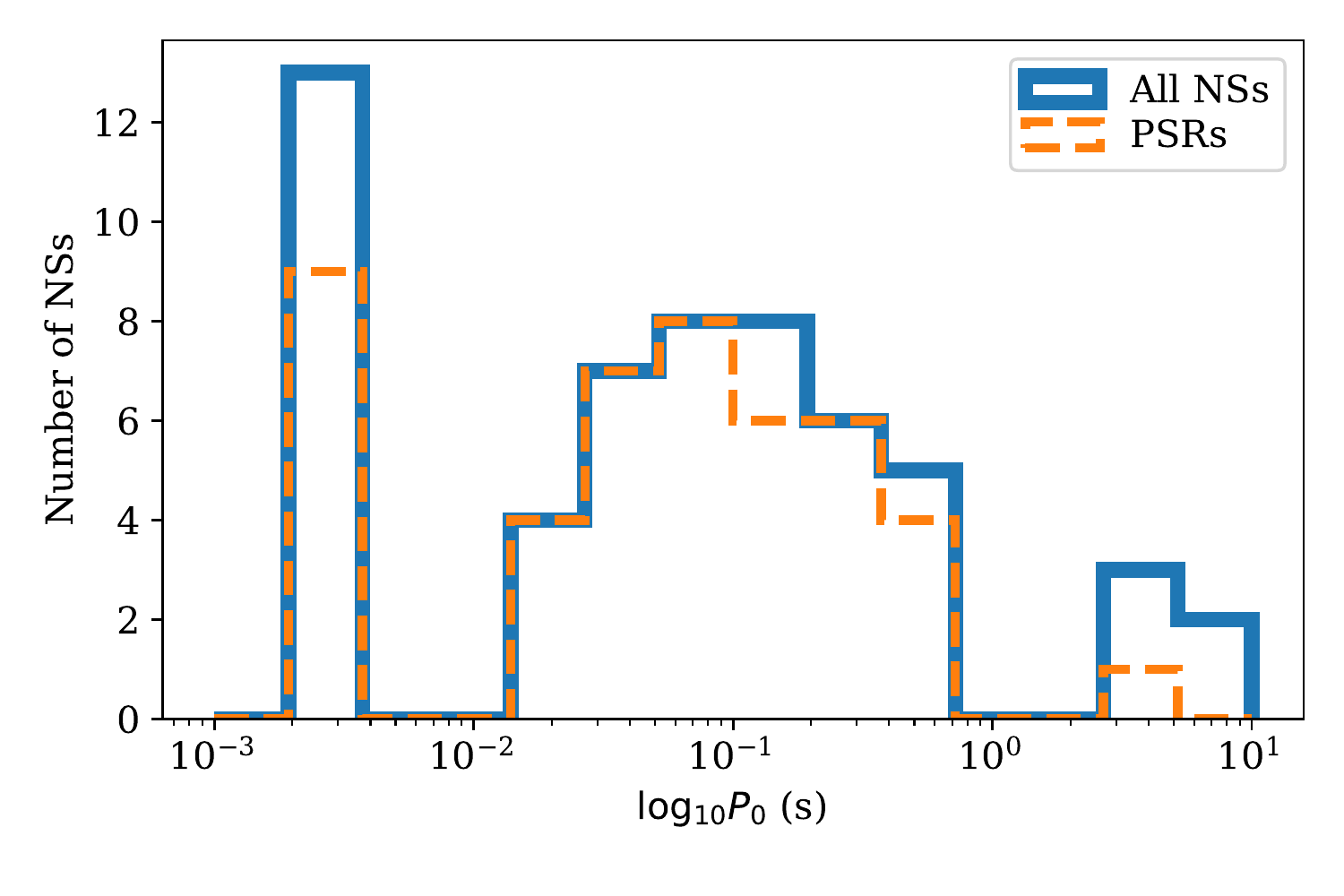}
    \caption{The histogram for computed initial periods of NSs associated to SNRs.  }
    \label{fig:P0}
\end{figure}

All computed initial periods are summarised in Table~\ref{t:catalogue}. We show the distribution of computed initial periods in Figure~\ref{fig:P0} using logarithmic bins. It is clear that three separate populations of NSs are present: (1) majority of NSs have computed initial periods ranging from $0.01$~s to 2~s, (2) NSs with artificially assigned periods all have $P = 0.002$~s clearly far below the computed periods for the majority of NSs, and (3) magnetars with computed initial periods ranging from 2~s to 10~s. 
Probably, magnetars' initial periods might be affected differently in comparison to normal radio pulsars by circumstances of a supernova explosion such as kicks, fallback discs, interaction with surrounding medium, etc. Thus magnetars' initial periods could have been  much shorter (see Discussion) and their computed initial periods are not representative of their actual initial periods. Therefore, in this section we only concentrate on the computed initial periods in the range 0.01--2~s for normal radio pulsar. It is our new selected sample and it includes 35 radio pulsars.

\begin{figure*}
    \begin{minipage}{0.49\linewidth}
	\includegraphics[width=\columnwidth]{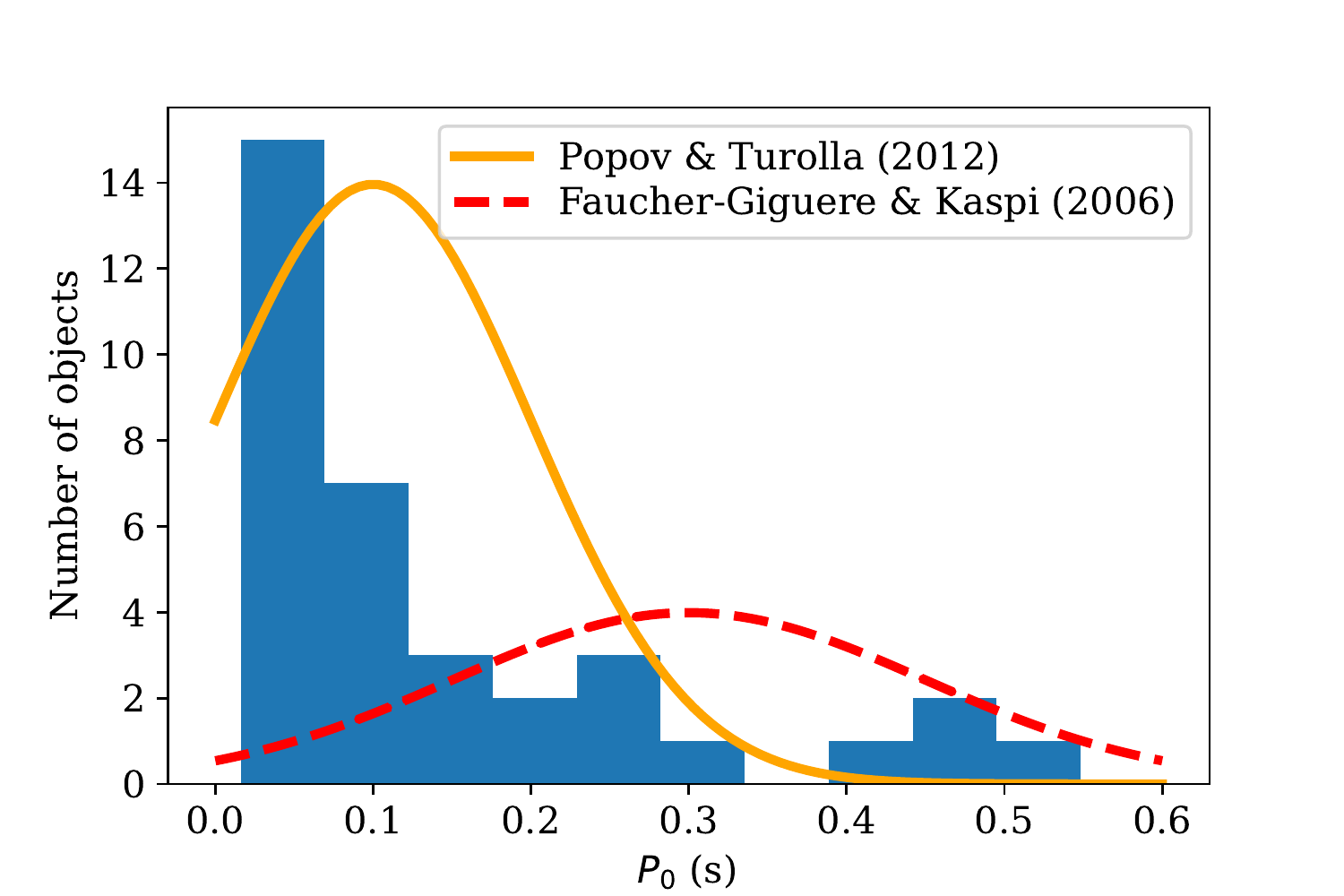}
	\end{minipage}
	\begin{minipage}{0.49\linewidth}
	\includegraphics[width=\columnwidth]{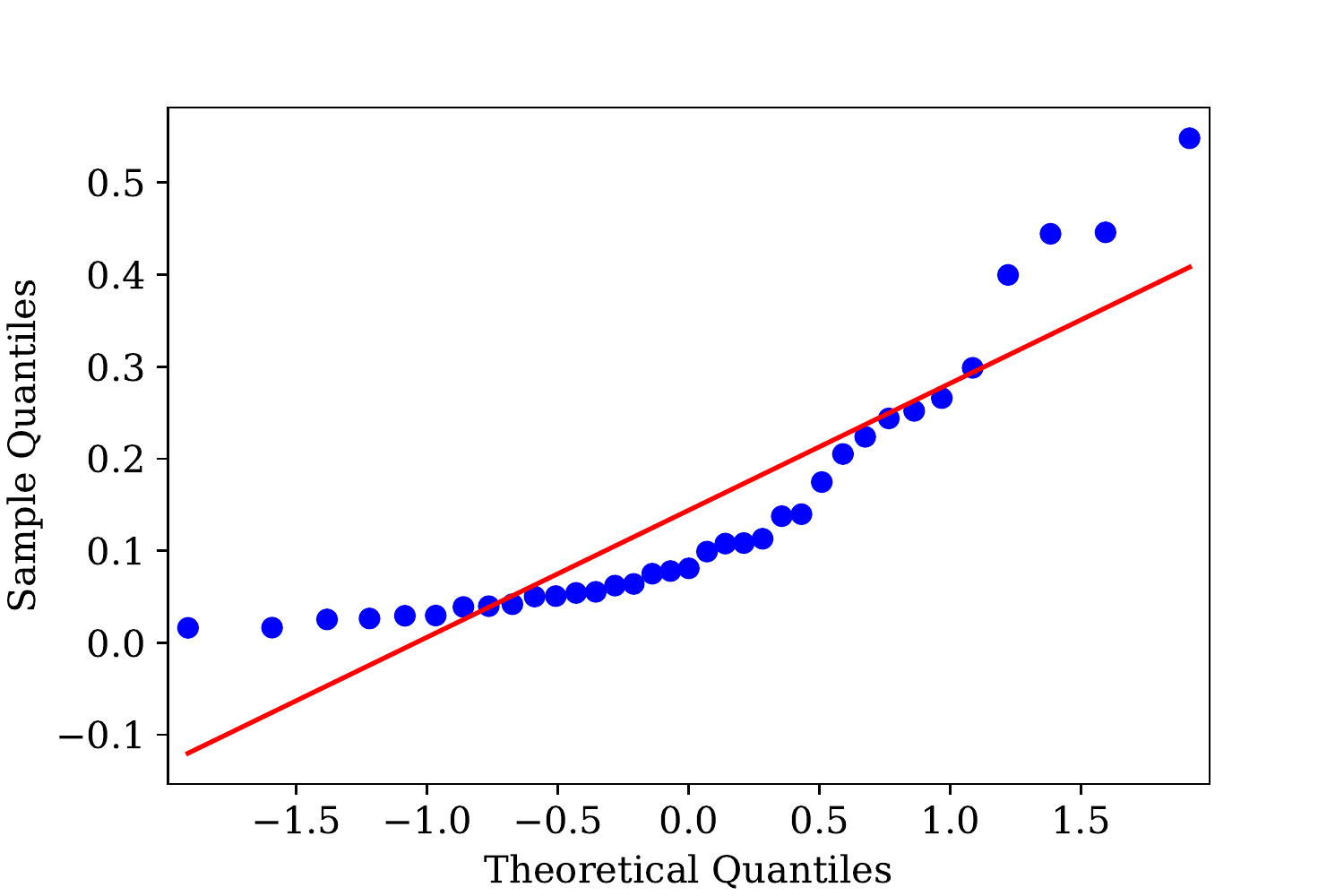}
	\end{minipage}
	\begin{minipage}{0.49\linewidth}
	\includegraphics[width=\columnwidth]{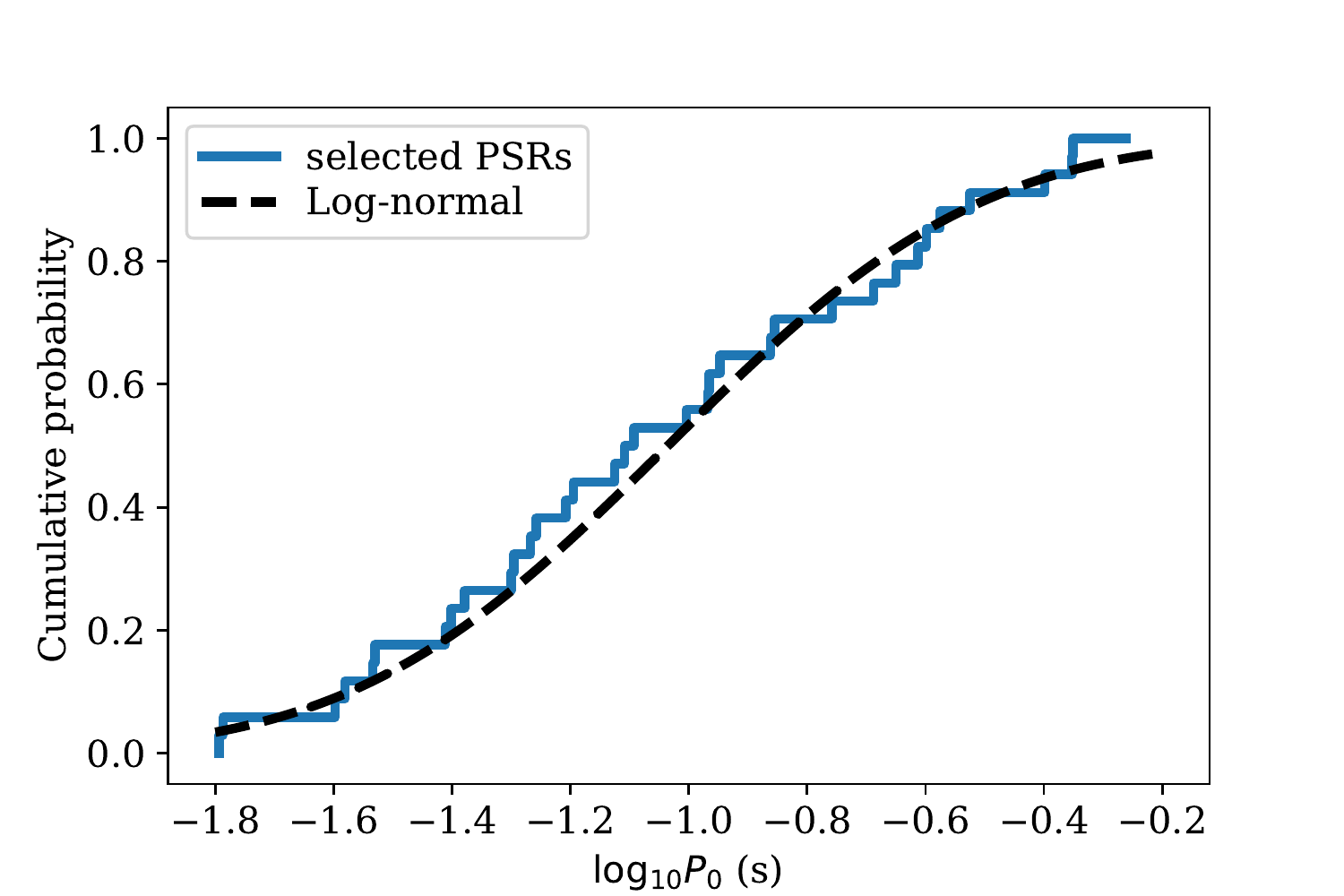}
	\end{minipage}
	\begin{minipage}{0.49\linewidth}
	\includegraphics[width=\columnwidth]{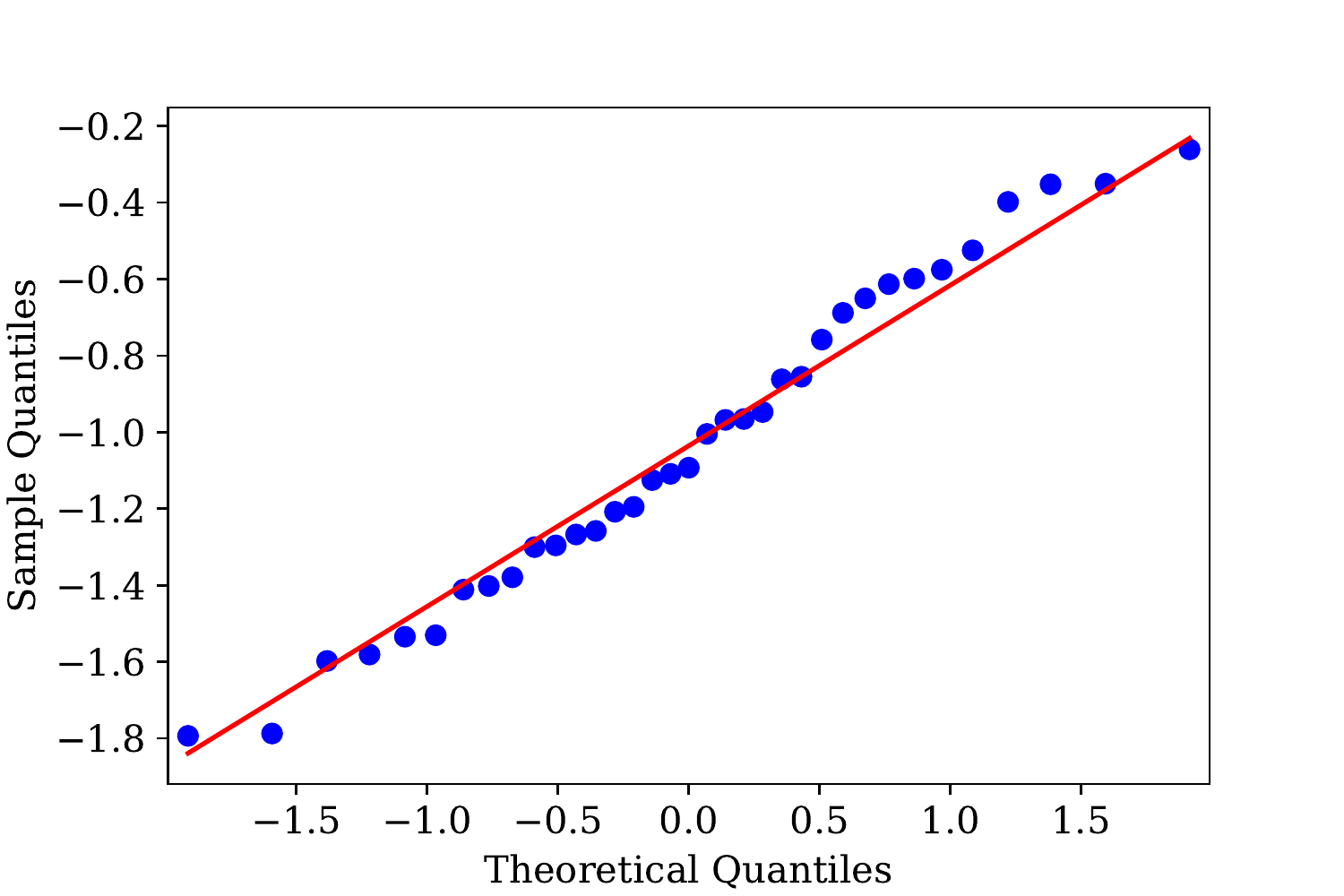}
	\end{minipage}
    \caption{Upper left panel: histogram for distribution of computed initial periods for selected radio pulsars. Upper right panel: Q-Q plot for normal distribution for computed initial periods of selected radio pulsars. Lower left panel: log-normal cumulative distribution for computed initial periods with $\mu_p = -0.96$ and $\sigma = 0.43$. Right lower panel: Q-Q plot for log-normal distribution for computed initial periods of selected radio pulsars. Red solid line is reference line. }
    \label{fig:logP0_distr}
\end{figure*}

We show the distribution of the computed initial periods for selected sample in Figure~\ref{fig:logP0_distr} (upper left panel). It is clear from this figure that the distribution is not uniform and not normal. We confirm our guess using the Shapiro test for normality ($p=3\times 10^{-5}$) and Q-Q plot, see Figure~\ref{fig:logP0_distr} (upper right panel). The mean value and the standard deviations are estimated $\mu_p = 0.14$~s and $\sigma = 0.14$, which is similar to results obtained by \cite{PopovTurolla2012}. Comparing the distribution suggested in that work with the histogram of computed periods, we see that this distribution provides an acceptable fit to the central mass of the data, but underestimates the number of pulsars with periods $\approx 0.5$~s. Another normal distribution ($\mu_p = 0.3$~s and $\sigma_p = 0.15$~s suggested by \citealt{FaucherGiguere2006ApJ}) underestimates number of pulsars with initial periods shorter than $0.1$~s. 

Instead of continuing with the above, we notice that the distribution of computed periods in logarithmic axis, see Figure~\ref{fig:P0} has a simple shape which resembles the normal distribution curve. Therefore, we find the base-10 logarithmic values for the periods and again performed the Shapiro test. It gives us $p$-value of $0.45$, which is not enough evidence to reject the hypothesis that the distribution is log-normal. We estimate the mean value of base-10 logarithm as -1.04 i.e. spin periods $p=0.091$~s and standard deviation as 0.42. We show the comparison of cumulative distribution and Q-Q plot in lower panels of Figure~\ref{fig:logP0_distr}. The shape of cumulative log-normal distribution is remarkably similar to histogram of base-10 logarithms of computed initial periods for PSRs.

We summarise results of our analysis for different samples in Table~\ref{tab:logB_res}. The log-normal distribution could successfully describe only the sample including radio pulsars. The hypothesis that log-normal distribution describes the distribution of initial periods for strongly magnetised NSs is excluded at 3~percent confidence level.

\subsubsection{Maximum likelihood estimate}
In this section, we introduce the maximum likelihood estimate for initial periods of NSs. In our case, the likelihood function contains the following four components: (1) the initial distribution of periods $\mathcal P (P_0, \mu_p, \sigma_p)$ with its parameters $\mu_p$ and $\sigma_p$; (2) distribution of actual pulsar ages $\mathcal P_t(t)$; (3) distribution of measured ages when actual age is given $\mathcal P_t(t' | t)$ and (4) distribution of measured periods and magnetic fields when initial period and actual age are given $\mathcal P(P', B_0 | P_0, t)$. The joint probability is a multiplication of all these four factors:
\begin{equation}
\mathcal P(P', B_0, t', t, P_0) = \mathcal P(P', B_0 | P_0, t) \mathcal P_t(t' | t) \mathcal P_t(t) \mathcal P_p(P_0, \mu_p, \sigma_p), 
\label{eq:joint}
\end{equation}
where $P'$ is the measured instantaneous period of
radio pulsars, $B_0$ is the effective measured magnetic field. In this section, we assume that this effective magnetic field is the same as at the moment of NS birth. As for the remaining variables: $t'$ is the measured age of SNR and $t$ is unknown actual age of NS. In this expression we also use unknown actual initial period of NS $P_0$. Different functions in eq. (\ref{eq:joint}) are described as the following. The conditional probability for pulsar initial period and magnetic field is a delta function:
\begin{equation}
\mathcal P(P', B_0 | P_0, t) =  \delta \left(P' - \sqrt{P_0^2 + 2 \kappa B_0^2 t }\right)      
\end{equation}
It allows us to restrict the parameter space and pair only $t$ and $P_0$ which correspond to measured pulsar period $P'$. 
The conditional probability for measured age given the actual age is a bounded uniform distribution:
\begin{equation}
\mathcal P_t(t'|t) = \left\{
\begin{array}{cc}
    1 / (b_t - a_t) & \mathrm{if} \; a_t\leq t'\leq b_t   \\
    0     & \mathrm{otherwise} 
\end{array}    
\right.
\end{equation}
where $a_t$ and $b_t$ are lower and upper bound for SNR age found in observations. In our analysis we also decide to check how sensitive our result is to the exact values of these bounds. That is why we additionally consider a case when $a_t = 0.5 t$ and $b_t = 1.5 t$. In this case, the age limits obtained in terms of $t'$ are $[0.66 t', 2 t']$. This approach assumes that $b_t / a_t  = 3$, which is quite similar to many real SNR age estimates, as can be seen from Table~\ref{t:catalogue}. However, in some cases the SNR lower age limit is more than order of magnitude smaller then the upper limit as in the case of CXOU J171405.7-381031 where SNR age estimate is [0.65, 16.8]~Kyr.

The initial distribution for pulsar ages is chosen to be uniform and it covers the whole range of possible ages when the SNR still can be associated with radio pulsars:
\begin{equation}
\mathcal P_t(t) = \left\{
\begin{array}{ccc}
1/t_\mathrm{max} & t< t_\mathrm{max} \\
0                & \mathrm{otherwise}
\end{array}
\right.
\end{equation}
The exact boundary $t_\mathrm{max}$ does not affect the result given that it is larger than any SNR upper age limit.  
As for the initial period distribution, we consider two options. First, we modify the normal distribution and write it in the following form:
\begin{equation}
\mathcal P_p (P_0, \mu_p, \sigma_p) = \frac{C}{\sqrt{2\pi}\sigma_p} \exp\left(-\frac{(P_0 - \mu_p)^2}{2\sigma_p^2}\right)    
\end{equation}
where the normalisation constant $C$ is:
\begin{equation}
C = \left(1 - \frac{1}{2} \left[ 1 + \mathrm{erf} \left( -\frac{\mu_p}{\sqrt{2} \sigma_p} \right) \right] \right)^{-1}  
\end{equation}
If $\mu_p / \sigma_p > 3$ the normalisation constant is $C\approx 1$. But the normal distribution seems to partially fit our data if $\mu_p / \sigma_p \approx 1$, which means that a significant part of the normal distribution is truncated (negative initial periods are impossible). Therefore, proper normalisation is crucial.

The log-normal distribution is:
\begin{equation}
\mathcal P_p (P_0, \mu_p, \sigma_p) = \frac{\log_{10} (e)}{P_0 \sigma_p \sqrt{2\pi}} \exp\left(-\frac{(\log_{10} (P_0) - \mu_p)^2}{2\sigma_p^2} \right)    
\end{equation}

At this point, we can write the functional form for the joint probability eq. (\ref{eq:joint}). This form, however, includes two unknown values: actual $P_0$ and actual age $t$. In order to get rid of these variables we integrate over them:
\begin{equation}
\mathcal P(P', B_0, t') = \iint\limits_{0\, 0}^{t_\mathrm{max}\, P} \mathcal P(P', B_0 | P_0, t) \mathcal P(t' | t) \mathcal P(t) \mathcal P_p(P_0) dP_0 dt
\label{eq:integ}
\end{equation}
where we omit $\mu_p$ and $\sigma_p$ parameters for conciseness. This procedure means that we include all acceptable values of $t$ and $P_0$ in our analysis for each NS. 
It is easy to analytically compute integral over $P_0$ using delta function. We compute $P_0$ as:
\begin{equation}
P_0^2 = P'^2 - 2 \kappa B_0^2 t   
\end{equation}
Computing the integral over the delta function, we make a substitution in the form $x = P' - \sqrt{P_0^2 + \kappa B_0^2 t}$, which results in appearance of additional factor:
\begin{equation}
\frac{P' dx}{\sqrt{P'^2 - 2\kappa B_0^2 t}} = dP_0    
\end{equation}
Therefore, the eq. (\ref{eq:integ}) becomes:
\begin{equation}
\mathcal P(P', B_0, t') = \int\limits_{a_t}^{b_t} \mathcal P_p\left(\sqrt{(P'^2 - 2\kappa B_0^2t)}\right) \frac{P' dt}{\sqrt{P'^2 - 2\kappa B_0^2 t}}     
\label{eq:ind_likel}
\end{equation}
where we removed some constant factors which do not affect the result of maximum likelihood calculations for the sake of conciseness. There is an additional hidden constraint related to the fact that not all ages $t$ are possible. We only integrate over the range where $P' - \sqrt{2\kappa B_0^2 t} > 0$.  
The integration limits change because we expand the uniform distributions and selected only the range where $\mathcal P(t'|t)$ is non-trivial.
This eq. (\ref{eq:ind_likel}) depends on $\mu_p$ and $\sigma_p$ and represents the likelihood for individual pulsars to have measured $P', B_0$ and $t'$, given the parameters of the initial period distribution $\mu_p, \sigma_p$. We compute the integral in eq. (\ref{eq:ind_likel}) numerically using Gaussian quadrature method with $n=100$ nodes.
To construct the total likelihood, we compute the logarithm of likelihoods for individual objects and sum them as follows:
\begin{equation}
L (\mu_p, \sigma_p) = -\sum\limits_{i=1}^N \log \mathcal P(P'_i, B_{0, i} t'_i | \mu_p, \sigma_p)    
\end{equation}
To find estimates for parameters $\mu_p^*$ and $\sigma_p^*$ we minimise this log-likelihood (maximise the total likelihood). 

In order to compare two different initial period distributions we find the best parameters for both models and then compute the Akaike information criterion (AIC) values:
\begin{equation}
\mathrm{AIC} = 2 L_\mathrm{A} (\mu_p^*, \sigma_p^*) - 2 L_\mathrm{B} (\mu_p^*, \sigma_p^*)   
\end{equation}
if the number of free parameters is the same in both models. Model A is $\exp(\mathrm{AIC} / 2)$ times more probable than model B. We test our developed maximum likelihood approach using synthetic data in Appendix~\ref{s:synthetic}.

\subsubsection{Results of maximum likelihood analysis}
\label{s:maxmim_likelihood_res}

\begin{figure*}
    \begin{minipage}{0.49\linewidth}
    \includegraphics[width=\columnwidth]{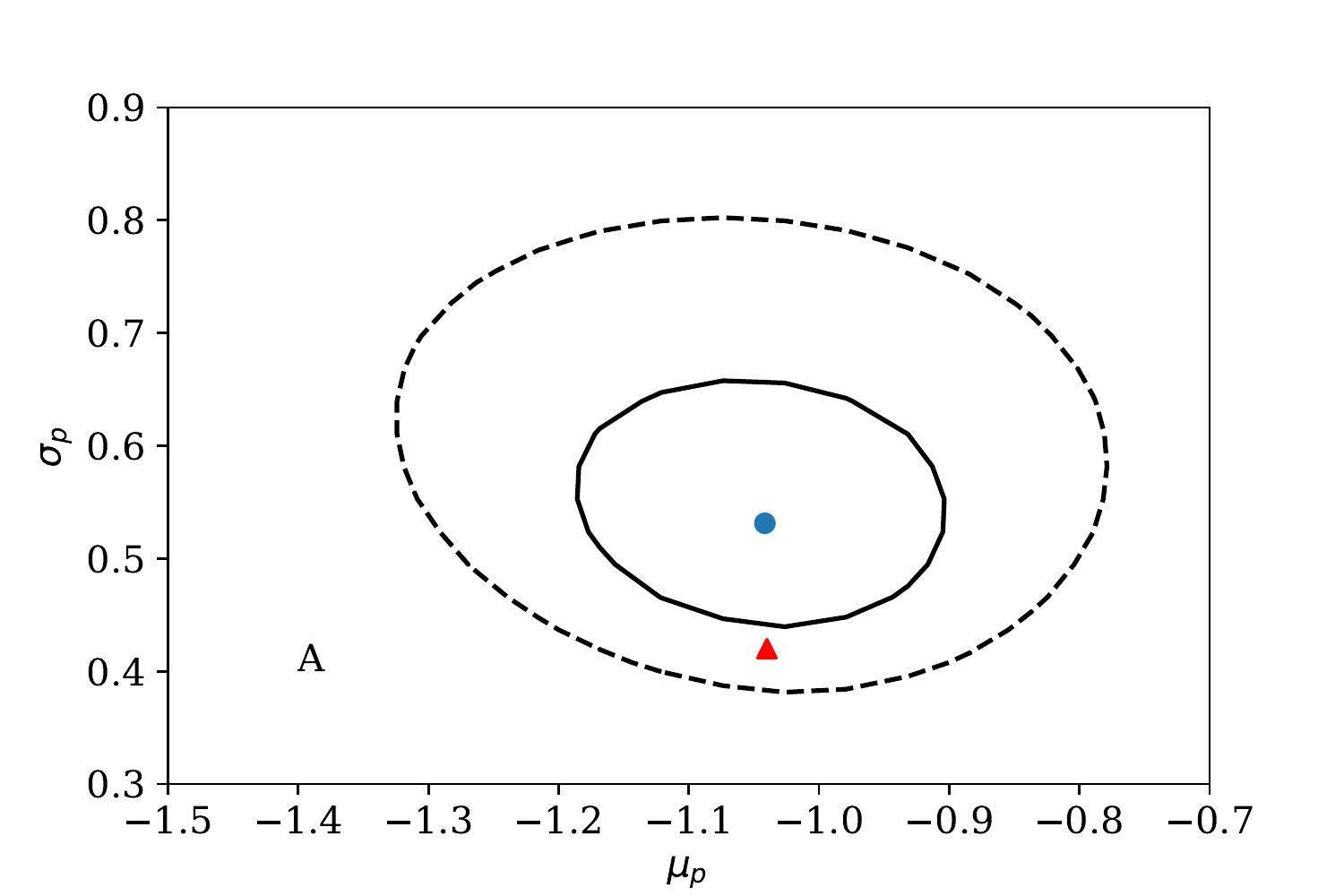}
    \end{minipage}
	\begin{minipage}{0.49\linewidth}
    \includegraphics[width=\columnwidth]{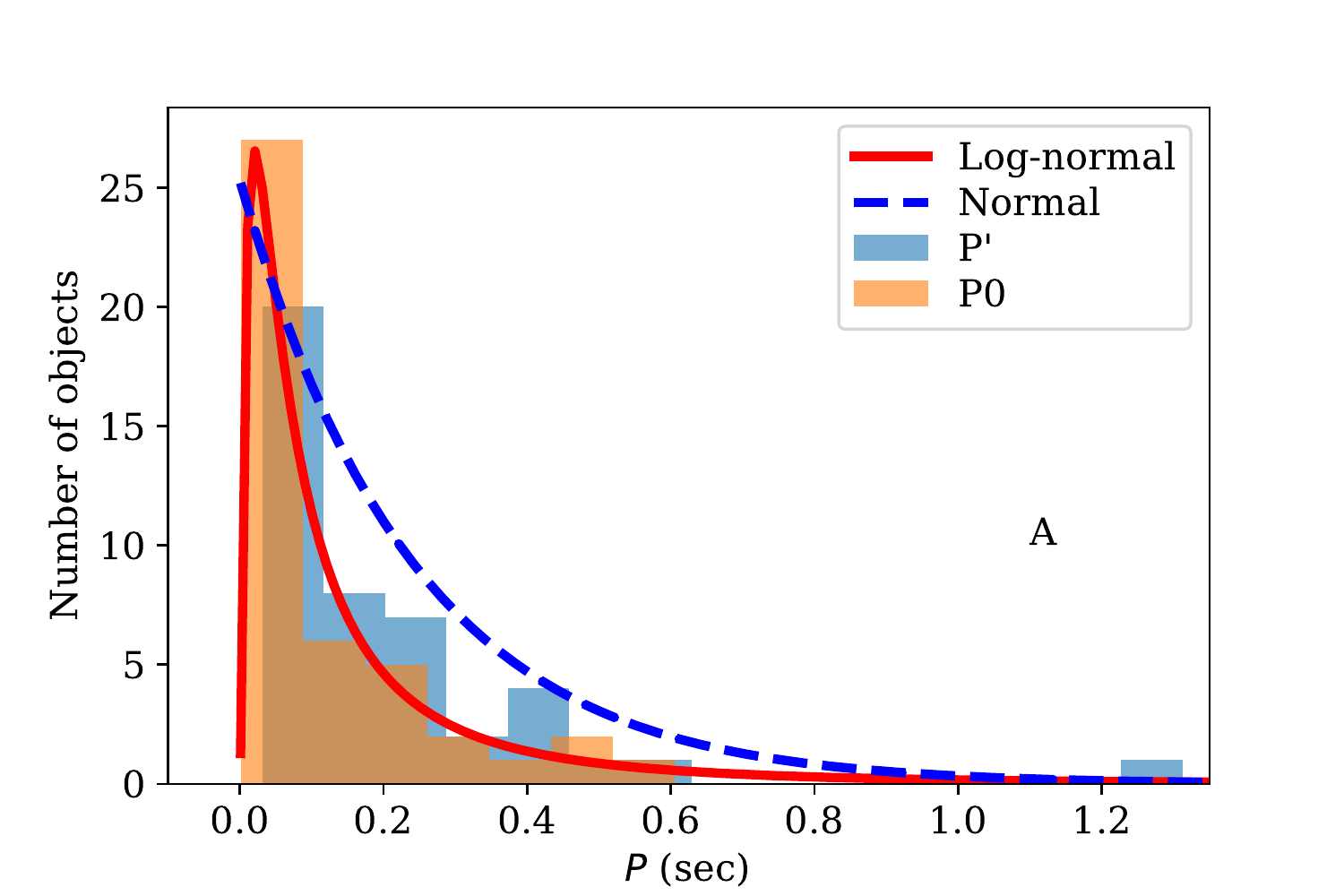}
    \end{minipage}
    \begin{minipage}{0.49\linewidth}
    \includegraphics[width=\columnwidth]{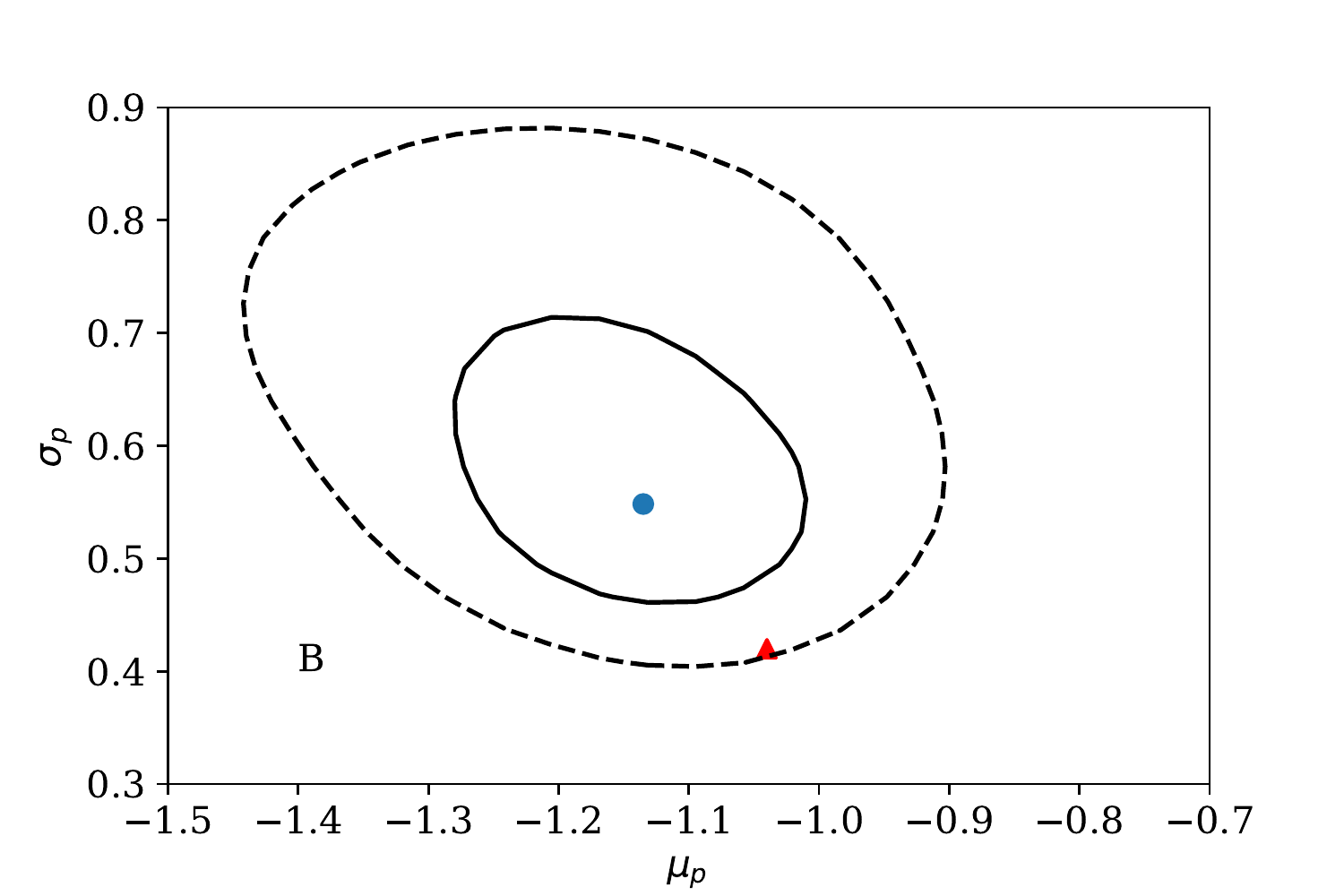}
    \end{minipage}
	\begin{minipage}{0.49\linewidth}
    \includegraphics[width=\columnwidth]{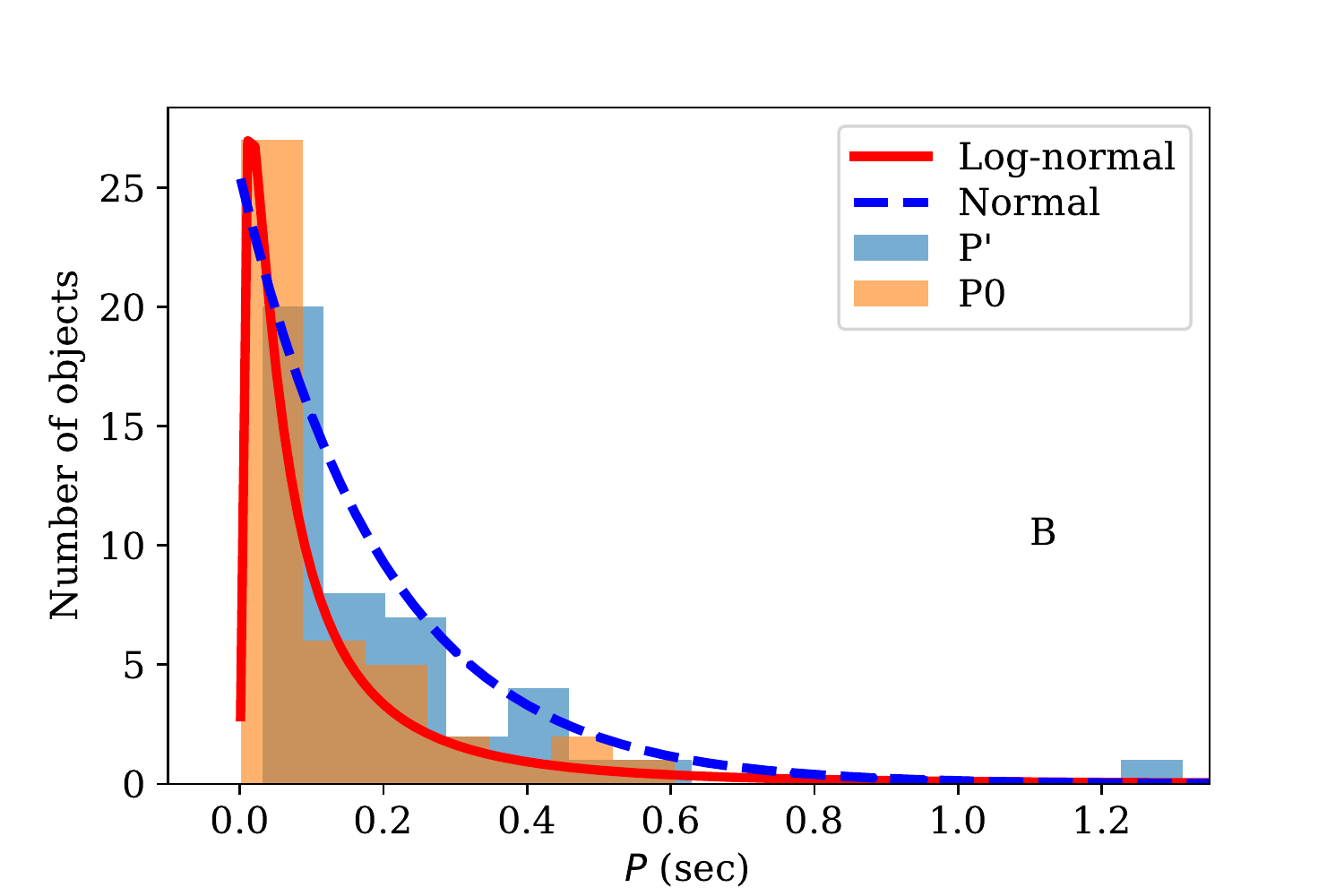}
    \end{minipage}
    \caption{Top panels: age ranges as in the catalogue; lower panels: $a_t = 0.5 t$ and $b_t = 1.5 t$. Left panels: contours of constant likelihood for the initial periods drawn from the log-normal distribution. Blue dot corresponds to the maximum likelihood, solid and dashed contours to 68 and 99 percent confidence intervals. Red triangle shows values found in the simplified analysis using only selected radio pulsars with estimated initial periods in range [0.01, 2]~s.  Right panels: histograms for measured periods of radio pulsars $P'$, their estimated initial periods $P_0$ (where possible), best fit log-normal (solid red curve) and normal distributions (dashed blue line).}
    \label{f:maximum_likel}
\end{figure*}

When we apply our maximum likelihood technique to catalogue data (45 radio pulsars excluding HBPSR) we obtain results summarised in Table~\ref{t:max_likel_res}. We illustrate these results in Figure~\ref{f:maximum_likel}. From this, it follows that the normal distribution does not describe initial periods of radio pulsars well enough. On the contrary, the log-normal distribution provides a much better model. In order to additionally check if the log-normal distribution with parameters in Table~\ref{t:max_likel_res} describes the data well, we draw 60 initial periods (comparable to initial size of our catalogue). Then we compare these synthetic initial periods with initial periods provided in Table~\ref{t:catalogue} by means of two-sample Kolmogorov-Smirnov test\footnote{Two sample Kolmogorov-Smirnov test is used to check if two samples are drawn from the same but unknown distribution. }.
We obtain $D = 0.183$ and $p=0.319$, which means that there is not enough evidence to reject hypothesis that they are drawn from the same distribution. We repeat the same procedure for 100 synthetic initial periods and Kolmogorov-Smirnov test provides us with the same conclusion that the synthetic periods and the initial periods can be described using the same log-normal distribution.

\begin{table}
    \centering
    \caption{Parameters estimated using the maximum likelihood technique. Assumption A stands for $a_t$ and $b_t$ same as the SNR ages and assumption B stands for the case $a_t = 0.5 t$ and $b_t = 1.5 t$. Confidence intervals are 68\% i.e. 1-$\sigma$ interval.}
    \label{t:max_likel_res}
    \begin{tabular}{rrcccc}
    \hline
    Assumption & Model       &  $\mu_p^*$        &  $\sigma_p^*$         & AIC \\
    \hline
    A & Log-normal  &  $-1.04_{-0.2}^{+0.15}$  & $0.53_{-0.08}^{+0.12}$ & ---   \\
      & Normal      &  $-10$      & $1.56$ & -18.1 \\
    \hline
    B & Log-normal  &  $-1.13\pm 0.13$  & $0.55_{-0.1}^{+0.15}$ & ---   \\
      & Normal      &  $-10$      & $1.48$ & -20.9 \\
    \hline
    \end{tabular}
\end{table}

When the algorithm tries to fit the normal distribution, it concentrates at the tail because of large periods present in the population. Therefore, it shifts $\mu_p$ far in the negative region and increases $\sigma_p$. Unfortunately, our procedure does not work stably with $\mu_p < -10$, making it is impossible to know whether the normal distribution provides a reasonable fit. In any case, the parameter space $\mu_p > 0$ is completely excluded. 

Results obtained using the simplified analysis for selected radio pulsars are within 99~percent confidence interval of those obtained with the maximum likelihood technique. It means that our estimates are reasonable. We seem to underestimate the width of the log-normal distribution in the simplified analysis.

\section{Short term magnetic field evolution}
\label{s:linear_model_B_evol}

\begin{figure}
	\includegraphics[width=\columnwidth]{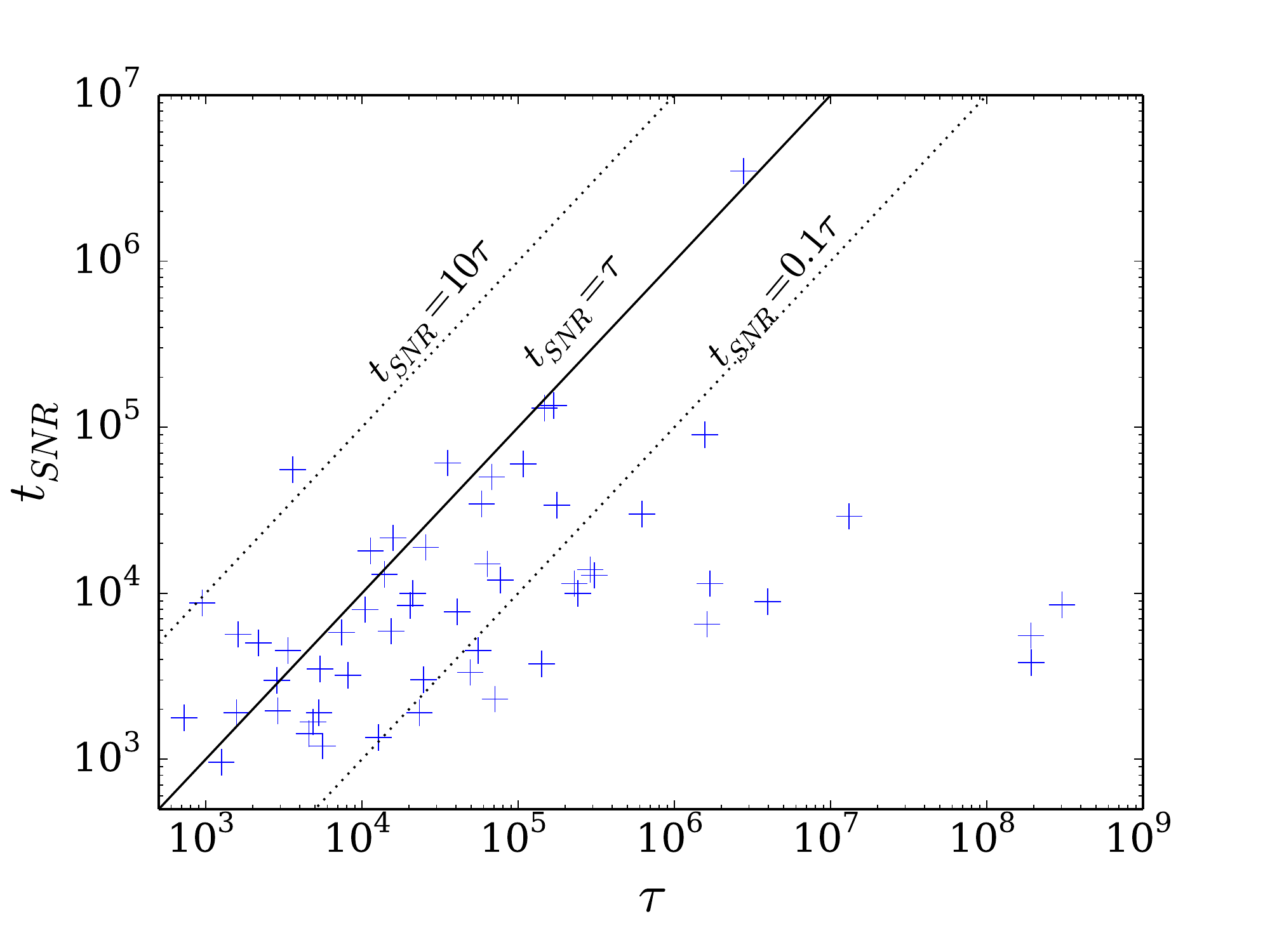}
    \caption{Characteristic age $\tau$ versus SNR age $t_{SNR}$ for the sources included in the sample.}
    \label{fig:tau_t}
\end{figure}

Comparing the characteristic age of pulsars with the age of the SNR that hosts it, we notice that for 37 sources the two ages lie within a factor of ten of each other, see Figure \ref{fig:tau_t}. Whereas regarding the remaining 19 sources, for 18 of them the characteristic age is at least a factor of ten larger than the characteristic age, and for only one source the SNR age is about factor of ten higher than the characteristic age. Overall, in 12 sources the characteristic age is smaller than the age of the SNR, whereas in the remaining 44, the age of the SNR is smaller than of the pulsar. A possible interpretation of this discrepancy in the ages could be attributed to magnetic field evolution.

\begin{figure}
	\includegraphics[width=\columnwidth]{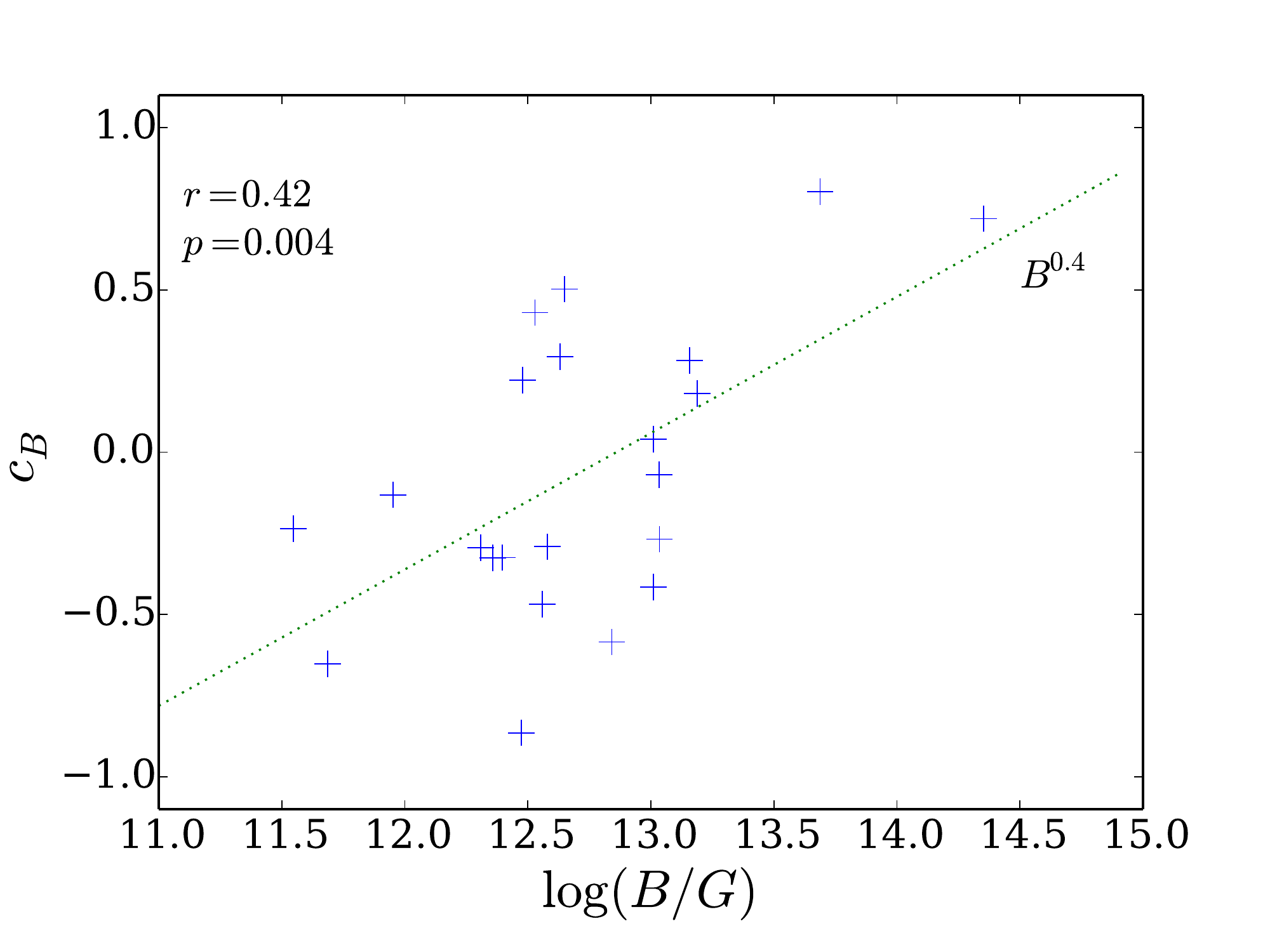}
    \caption{Magnetic field linear change coefficient $c_B$ versus the logarithm of the magnetic field. }
    \label{fig:B_Cb}
\end{figure}
\begin{figure}
	\includegraphics[width=\columnwidth]{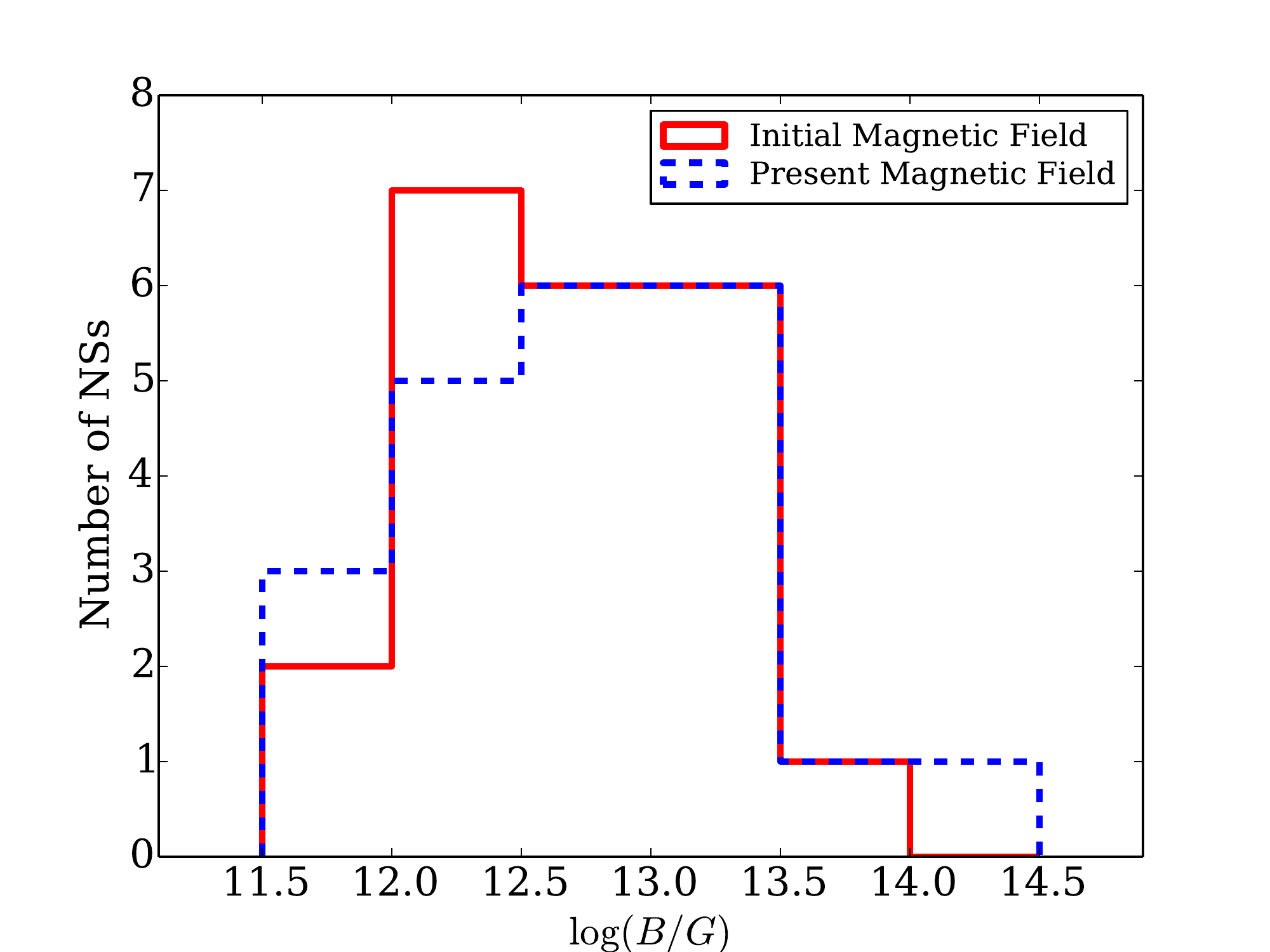}
    \caption{Histogram of the current and the initial magnetic field obtained by the short-term evolution model. }
    \label{fig:B_distr_initial_current}
\end{figure}

The spin-down dipole model assumes that the properties of a pulsar such as the strength of the magnetic field and its moment of inertia do not change with time. As we discussed above, NS characteristic ages estimated with this model i.e. eq. (\ref{eq:tau_ch_age}) in some cases are inconsistent with SNR ages estimated using the SNR expansion rate. This discrepancy might be explained if we assume that the strength of poloidal dipolar magnetic field changes with time as it was proposed in several models. While the long term trend supports that the magnetic field of NS decays due to the activity of Ohmic dissipation \citep{2007PhRvL..98g1101P,2015PhRvL.114c1102H}, it is possible that temporary growth of the magnetic field dipole component may have occurred in some neutron stars. This could be due to internal magnetic field reconfiguration \citep{2015MNRAS.446.1121G} or the reemergence of a buried magnetic field \citep{2011MNRAS.414.2567H,2016MNRAS.462.3689I,2020MNRAS.495.1692G,2021ApJ...909..101I}. Therefore, we model the evolution of the dipole magnetic field through a linear expression given by the following equation:
\begin{equation}
\label{td}
    B(t)=B'\left(1+c_B \frac{t-t_{SNR}}{t_{SNR}}\right)\,,
\end{equation}
where $B'$ is the present magnetic field,  $c_B$ is a constant of proportionality, that can be either positive or negative and indicates the increase or decrease of the magnetic field respectively, and $t_{SNR}$ is the SNR’s age. The underlying assumption for the above relation is that the SNR age represents the actual age of the NS, and any deviation between the two ages is due to magnetic field evolution.  While the linear expression assumed here is a simplification, the majority of these sources are young. Thus, even more complicated evolutionary profiles can be approximated by a linear model for short time-scales up to a few $10^4$yr compared to older NS whose ages exceed $10^{6}$yr. Essentially, we expand the unknown magnetic field evolution function into its Taylor series:
\begin{equation}
B(t) = B_0 + B' (0) t + \frac{B'' (0)}{2} t^2 + O (t^3).     
\end{equation}
And we include in our estimate only two first terms.

We substitute the expression for the magnetic field (\ref{td}) into equation (\ref{eq:bp}) and integrate it with respect to time, assuming that the initial period is $0$ and the current period is the measured one. This yields an equation from which can determine $c_B$. While formally we can apply this to all 56 pulsars of our sample, the results can have some physical significance only for the sources whose characteristic and SNR ages do not differ much. For instance, for systems where the value of $c_B$ is $1$, the initial magnetic field is zero and similarly for systems where $c_B$ is smaller than $-1$, it implies that within a NS's lifetime the magnetic field has decayed  by 50~percent. Here we focus on systems with $-1<c_B<1$. We find that there are 22 such systems, in 9 $c_B$ is positive, implying that the magnetic field has grown, whereas in 13 systems $c_B$ is negative suggesting an overall decay, as in Figure~\ref{fig:B_Cb}. 

We find that there is a correlation between the magnetic field strength and $c_B$, with Pearson correlation coefficient $r=0.42$ and corresponding $p$-value $p=0.004$. 
Thus, we can reject the hypothesis that $c_B$ and $\log_{10} B$  are independent at significance $0.1$~percent. Nevertheless, a linear relation  between $c_B$ and $\log_{10} B$ is mild as the Pearson coefficient is quite small.

We can explore the relation between $c_B$ and $\log_{10} B$ by performing a linear regression. We find a scaling of the form $c_B\propto B^{0.42\pm 0.13}$.  The linear regression line crosses $c_B = 0$ at $10^{12.85\pm 1.67}$~G, which means that NSs with a magnetic field stronger than $10^{12.85}$G seem to have mostly undergone a phase of growth, whereas the ones with weaker field seem to have suffered a decay. 
Using the values found above we can directly evaluate the initial magnetic field of these pulsars by setting $t=0$ in equation (\ref{td}). These results are shown in Figure \ref{fig:B_distr_initial_current}. The base-10 logarithmic average of the present magnetic field of the pulsars included in this sample\footnote{This is a part of the original sample selected according to specific rules, which is why its mean and standard deviation differ from ones summarised in Table~\ref{tab:logB_res}} is $12.70$, with a standard deviation of $0.60$. The corresponding initial field has a base-10 logarithmic average value of $12.68$ and a standard deviation of $0.48$. This is indicative of the behaviour found, where sources with higher dipole magnetic fields have undergone a growth phase, whereas the ones with weaker magnetic dipole fields have undergone an decrease phase.

We report that there is no statistically significant correlation between the age (either characteristic or SNR) and $c_B$.

\section{Discussion}
\label{DISCUSSION}

\subsection{Initial periods of magnetars}
It was suggested in the literature that magnetars slow down much more efficiently at earlier stages of the NS evolution, see e.g. \citep{2004ApJ...611..380T} and references therein. These authors show that most of magnetar rotational energy can be extracted during the first 10~sec of their evolution. 
Therefore, their subsequent rotational evolution is practically irrelevant, as the bulk of spin-down has occurred early in their lives.

It was also suggested in the literature that strong dipolar fields of magnetars could be formed due to their extreme fast rotation, see e.g. \citep{DuncanThompson1992ApJ, Raynaud2020SciA} and references therein. The initial periods has to be shorter than $\approx 6$~msec in this case.  
It is interesting to estimate the fraction of NS born with very short rotational periods using the log-normal distribution for initial periods found in Section~\ref{s:maxmim_likelihood_res}. The fraction of NSs born with $P_0 < 5$~ms is 0.0087 which is much less than the fraction of magnetars. Thus, our sample of young NS contains 8 magnetars out of 68 NSs, which is a fraction of $\approx 0.118$. 
Therefore, true NS initial periods might be even shorter than ones estimated here. Alternatively magnetars could be form from a sub-population of progenitors with special properties e.g. rotation rate.

\subsection{Effect of $n\neq 3$}
\label{s:non3n}
In this section, we check if our analysis of initial pulsar periods depends significantly on assumption of $n=3$. In general, a pulsar slows down as follows:
\begin{equation}
P^{n-1} \frac{d P}{dt} = K    
\end{equation}
This equation is a direct consequence of eq. (\ref{eq:mag-dip}), where we combine all constant factors in the value $K$. This equation has the following solution for $n\geq 2$:
\begin{equation}
P_0 = \left(P^{n-1} - (n-1) P^{n-2} \dot P t\right)^{1/(n-1)}    
\end{equation}
where $P_0$ is the initial period, $t$ is the pulsar age, and $P$ and $\dot P$ are currently measured period and period derivatives.
In the special case of $n=1$ the solution is:
\begin{equation}
\log (P_0) = \left(\log(P) - \frac{t}{2\tau}\right)    
\end{equation}

\begin{figure}
    \includegraphics[width=\columnwidth]{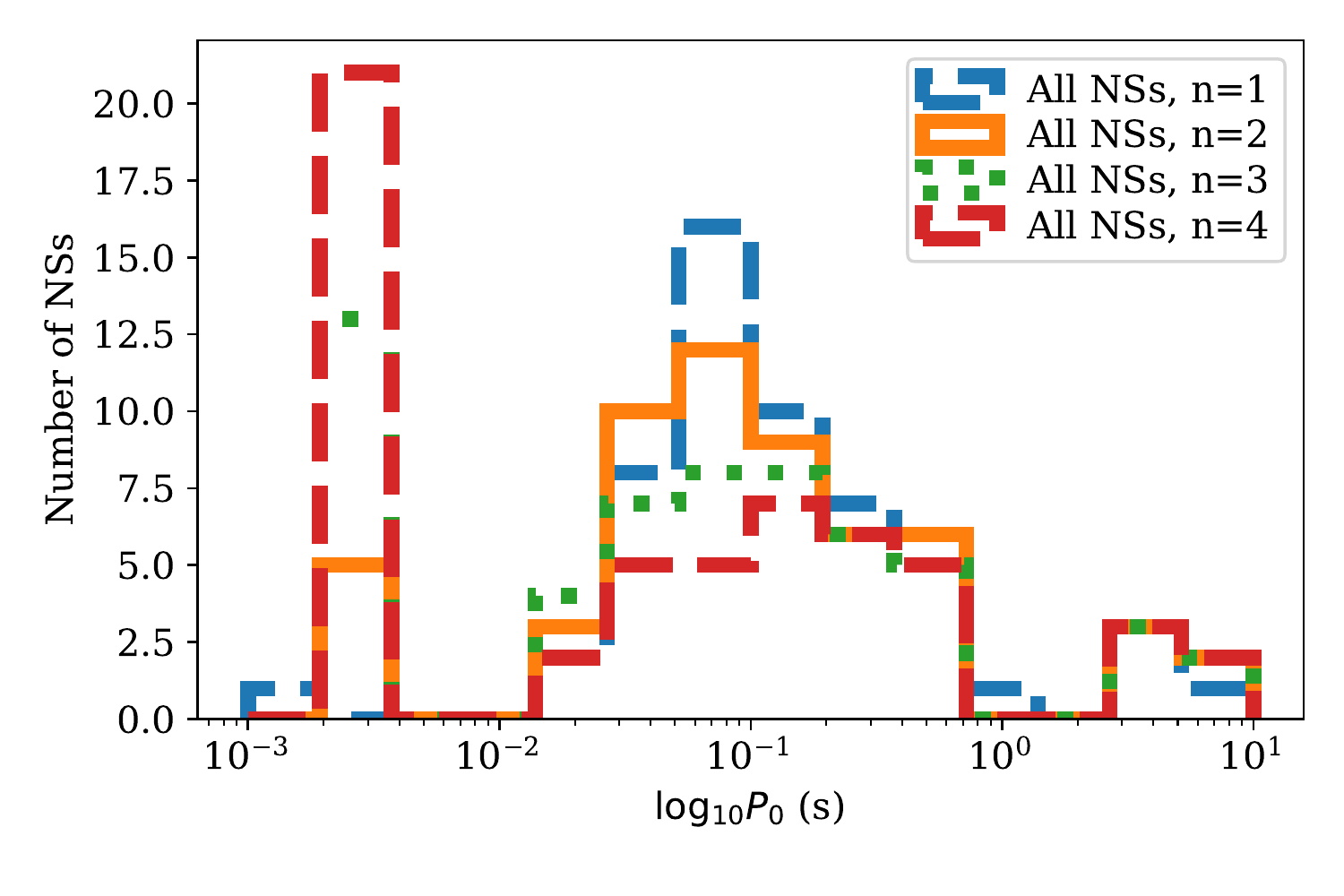}
    \caption{The histogram for computed initial periods of NSs associated to SNRs for different values of braking index.}
    \label{fig:P0_n}
\end{figure}

We compute the initial periods (where possible) using these new equations and plot these values in Figure~\ref{fig:P0_n}. Further, we select only initial periods in range from 0.01~s to 2~s. We include all neutrons stars which satisfy this period range. 

For all braking indices, the restored distribution looks quite similar. We also check if these computed values could be drawn from the normal distribution using the Shapiro test and find the following $p$-values: $<10^{-8}$ for $n=1$, $10^{-6}$ for $n= 2$ and $8\times 10^{-4}$ for $n=4$. Thus, in all these cases we reject the hypothesis that initial periods are drawn from a normal distribution. In all these cases we also compute logarithms of periods and perform the Shapiro test for these values. We obtain the following $p$-values: $0.19$ for $n=1$, $0.13$ for $n=2$ and $0.52$ for $n=4$. Thus, there is not enough evidence to reject the hypothesis that computed initial periods are drawn from the log-normal distribution. 

We summarise the values of mean and standard deviations for base-10 logarithms of computed initial periods in Table~\ref{tab:logP0n_res}. All these values are very similar to the case of $n=3$. Thus the results of our analysis are not sensitive to the exact value of the braking index.

\begin{table}
    \centering
    \caption{Results of analysis of initial periods assuming different braking indices. In all cases $p$-value corresponds to log-normal distribution.}
    \label{tab:logP0n_res}
    \begin{tabular}{lcccc}
    \hline
    \multicolumn{5}{c}{Initial spin periods} \\
    \hline
    Braking index & N & $\log_{10} P$ & $\sigma_p$ & $p$-value \\
    \hline
    $n=1$          & 50 & -0.97        & 0.39     & 0.19 \\
    $n=2$          & 46 & -1.02        & 0.40     & 0.13 \\
    $n=3$          & 38 & -1.01        & 0.42     & 0.42 \\
    $n=4$          & 30 & -0.93        & 0.40     & 0.52 \\ 
    \hline
    \end{tabular}
\end{table}

\subsection{Effects of observational selection}
\label{s:selection}

In this study, we do not account for any selection effects, assuming that the sample we use is unbiased. 
However, this is a simplification, inevitable due to complex nature of different, sometimes concurrent, selection effects related to SNR, PSRs, and CCOs. In the case of NSs in SNRs, additional specific difficulties with biases appear because many sources are not found in large uniform surveys, but in dedicated observations of particular objects and/or regions. 
 
Selection effects related to SNRs are discussed in several papers, see e.g. \cite{1995MNRAS.277.1243G, 2019JApA...40...36G} and references therein. Generally speaking, to be successfully identified a SNR might have significant surface brightness and have angular size larger than some critical value determined by the angular resolution of a survey. 

SNRs diffuse as they expand and are less likely to be visible once they become older. In practice, most sources eventually merge with the interstellar medium at the age of $30$ kyrs \citep{2008ARA&A..46...89R}. This sets a limit on age of the sources comprising our sample and a selection of younger sources that would be more easily identified. Indeed, median age in our sample is $7.7$ kyrs. Due to this, younger sources are more likely to be included in the sample (except very young and distant objects which can have a small angular size). On the contrary, characteristic ages of the pulsars in the sample span a much wider range with the median age being $25$ kyrs. 

Surface brightness depends not only on the age of the SNR and properties of the surrounding medium, but also on distance to the observer. Thus, near-by sources can avoid identification due to large angular size (and so, low surface brightness).

Many selection effects are known for PSRs. They are often discussed in papers containing methodology or/and results of surveys and in population synthesis studies (see, e.g. \citealt{FaucherGiguere2006ApJ, 2018ApJ...861...44P}). We mention and discuss just two of them. Both make discovery of slower PSRs (i.e. born with longer initial periods and/or more magnetised) less favourable.

Firstly, slowly rotating PSRs become harder to discover due to the spin-beaming correlation. It is known that PSRs with longer spins have narrow beams \citep{1998MNRAS.298..625T}. This makes detection of a PSR with a longer period less probable. According to \cite{1998MNRAS.298..625T} (see their Fig.~10) it is roughly twice less probable to find a PSR with $P=0.5$~s than a PSR with $P=0.1$~s.
Evolutionary effect of this correlation might be not very significant for our analysis as we are dealing with relatively young objects. Typically, during $\sim $~few tens of kyr of evolution a normal radio pulsar cannot significantly decrease the width of the beam. But the situation is different if we consider a wider range of initial spins.
If NSs can be born with periods up to few tens of a second, then those with longer spins can avoid detection. So, potentially, the initial spin distribution can be somehow wider than obtained above.

Second, the probability to detect a PSR depends on its radio luminosity, $L_\mathrm{r}$. How this quantity is related to $P$ and $\dot P$ is uncertain. In the model favoured by \cite{FaucherGiguere2006ApJ} it is expected that PSRs with longer periods and smaller $\dot P$ have lower luminosities, and thus can avoid detection with higher probability. However, some authors (e.g. \citealt{2014ApJ...784...59S})  object this conclusion. Still, we can expect that on average older PSRs with longer initial spin periods can be underrepresented in SNRs due to lower $L_\mathrm{r}$.

Finally, we have to note that for most of the known CCOs periods and period derivatives are not measured. Most probably, they would correspond to low values of the effective magnetic field. Thus, the number of low-field objects in the total sample is underestimated.

\subsection{Alike pulsars with and without SNRs}
\label{s:alikepsr}

Determination of initial parameters of NSs can significantly influence understanding of important details of their origin. 
Recently, \cite{2021MNRAS.508..279C} analysed a large sample of PSRs in SNRs and compared it with a sample of short period PSRs ($P<0.5$~s) not associated with SNRs. 
These authors attribute difference between the two samples to the existence of two types of SN: low-energy and high-energy. The first SN type -- related to lower mass stars dying in e$^-$-capture explosions\footnote{also known as electron capture supernova explosion (ecSN)}, -- produces SNRs with shorter lifetime. Thus, PSRs born after such SNae have less probability to be observed inside a remnant. In this subsection, we discuss this hypothesis in relation  to the approach used above and conclusions of our study. 

Let us focus, at first, on the case of PSRs with spin-down ages $\lesssim10$~--~15~kyrs. There exists comparable number of objects with and without SNRs, both sharing quite similar values of $P$ and $\dot P$ (see Fig.~1 in \citealt{2021MNRAS.508..279C}, also our Fig.~\ref{fig:pdotp}).
If these PSRs have different origins (and so, presumably, different initial properties and, may be, even evolution) this would have influence on the approach used in our study, where all PSRs are treated within the same model.

It is worth to note that the hypothesis proposed by \cite{2021MNRAS.508..279C} can be criticised from two sides. On one side, the expected fraction of ecSN is not high enough to explain the formation of majority of PSRs without SNRs in the sample by \cite{2021MNRAS.508..279C}. For isolated progenitors this fraction is expected to be $\approx 4$\% \citep{2008ApJ...675..614P}. For binary progenitors the fraction can be significantly higher \citep{2017ApJ...850..197P} but below $\sim20$\%. On another side, it is expected that e$^-$-capture SN imparts low natal kicks \citep{2004ApJ...612.1044P}. In this case  many NSs produced in this formation channel are expected to stay members of binary systems (see e.g. \citealt{Igoshev2021MNRASChruslinska}), which is not the case for PSRs analysed by \cite{2021MNRAS.508..279C}. Therefore, ecSN can be responsible just for a very small fraction of isolated PSRs (with a low spatial velocity as a characteristic feature), which is not sufficient to explain all young PSRs without SNRs. 

Alternatively, if we observe two PSRs with similar $P$ and $\dot P$ -- one is within a SNR and another has no associated remnant, -- then it is quite possible that these two NSs simply have significantly different ages. There are several explanations for such an outcome, which we describe below.

As SNRs are relatively short-living objects, then even an age difference about few tens of kyrs can be crucial. Radio pulsars born with different initial periods reach the same observed periods at different ages.
Thus, two pulsars with $P\approx 0.3$~s and $B\sim10^{12}$~--~$10^{13}$~G (i.e., their spin-down ages are the same) can have real ages different by $\gtrsim 10^5$~yrs, if one has $P_0\approx P$, and another $P_0\ll P$.
In \cite{2021MNRAS.508..279C} the authors do not account for this possibility.

Effectively, apparent large values of $P_0$ can be a consequence of decaying magnetic field. Two NSs can have very different magnetic field history if they have different impurity parameters $Q$ and/or thermal evolution (e.g., due to different masses).
In addition, field re-emergence {or reconfiguration} can play a role. If for a NS the present-day external dipolar field value is higher than during the precedent evolution, then the spin-down age is shorter than the actual age. Thus, a SNR could already disperse. 
Without additional information it would be impossible to distinguish between different variants of the field evolution by simply analysing the present day values of $P$ and $\dot P$. 
 
 Different rates of spin-down can also result from different values of the initial magnetic inclination angle, $\chi_0$, or field topology. 
 All these possibilities (some of them are not analysed  in our study, also) can lead to an incorrect estimate of real ages of NSs based on present-day $P, \dot P$ values. 
Thus, it is important to obtain additional independent age estimates. Of course, the best way is to derive a SNR age (see e.g.,  \citealt{Suzuki2021ApJ} and references therein). 
Fig.~\ref{fig:tau_t} above clearly demonstrates how significantly a SNR age can be different from the characteristic age of the associated PSR. However, for PSRs not associated to any remnant, independent age estimates could be based on their kinematics or their thermal properties. Without independent age estimates in hand, conclusions about different origin of PSRs associated/non-associated to SNRs can be premature. 

Results presented in this paper indicate that the majority of observed  young pulsars can be described as a unique population, including objects in SNRs. However, this topic requires more attention.

\section{Conclusions}
\label{CONCLUSION}

We matched the ATNF catalogue of radio pulsars against the catalogue of supernova remnants and found 68 possible associations. Many of these pairs were known already. If we have multiple candidates for association between SNR and NS, we select only one pair where SNR age and NS characteristic age are the most similar.

We analysed the distribution over magnetic fields and spin periods for these young NSs. We found that the distribution for magnetic field could be successfully described with the log-normal distribution with parameters $\mu_B = 12.44$ and $\sigma_B = 0.44$ being slightly smaller in comparison to the result by \cite{FaucherGiguere2006ApJ}. They found $\mu_B = 12.56$ and $\sigma_B = 0.55$. This might be related to difference in assumptions, since \cite{FaucherGiguere2006ApJ} assumed no magnetic field evolution on timescales $\sim 10^8$~--~$10^9$~years while we only require constant magnetic field during first $10^5$~years. Moreover, \cite{FaucherGiguere2006ApJ} compared their results with all radio pulsars including one which could be potentially activated later on in their evolution like magnetars and central compact objects.
In comparison to \cite{PopovPons2010MNRAS}, who found for the initial magnetic field distribution $\mu_B = 13.25$, we did not include the magnetars. Our analysis including magnetars and CCOs gives $\mu_B = 12.57$ and $\sigma_B = 0.86$, but a hypothesis that distribution is log-normal is rejected at significance level of 3\%. Thus strongly magnetised NSs could be members of a different population as was discussed by \cite{Gullon2015MNRAS}. 

Contrary to previous studies of the initial period distribution for radio pulsars we found that this distribution cannot be successfully described using the normal distribution. We independently confirm this conclusion using a novel maximum likelihood method.
Instead, we suggested a log-normal distribution with $\mu_p = -1.04_{-0.2}^{+0.15}$ i.e. $P_0 \approx 0.09$~s and $\sigma_p = 0.53_{-0.08}^{+0.12}$ (68 percent confidence interval) based on analysis of 45 radio pulsars. This results agree reasonably well with simple analysis of computed initial periods for radio pulsars. Magnetars' period distribution (including the initial periods) seem to be a distinct from period distribution of isolated radio pulsars. Magnetars' periods concentrate at $P_0 > 2$~s. 

For the sources whose characteristic and SNR ages do not differ by more than a few times we can provide a linear model of magnetic field growth or decay. We have found that this process leads to a population whose logarithmic average magnetic field does not evolve significantly with time, yet, the standard deviation of the distribution increases as neutron stars age.

\section*{Acknowledgements}

 Work of A.I.P. is supported by STFC grant no.\ ST/W000873/1.
SP acknowledges support from the Russian Science Foundation, grant 21-12-00141. We thank anonymous referee for their comments. A.I.P. also thanks Dr. Alexander Mushtukov for useful discussion. KNG acknowledges funding from grant FK 81641, University of Patras ELKE.

\section*{Data Availability}

All data files and scripts in form of jupyter-notebook files are available at github: \url{https://github.com/ignotur/PSR_in_SNR} 
 



\bibliographystyle{mnras}
\bibliography{example} 




\appendix

\section{Catalogue of NS associated to SNR}
\label{appendix1}

\begin{table*}
\centering
\caption{Catalogue of NSs used in our analysis. }
\label{t:catalogue}
\begin{tabular}{cllcccrrcccrrc}
\hline
\# & NS & SNR & Type & Included & $B$ & $P$ & $\dot P$ & $P_0$ & $t_\mathrm{SNR}$ range & $\tau$  \\
   &    &     &       & Y/N      & (G) & (s) & (s/s)    & (s)   &  (Kyr)                   & (Kyr)   \\
\hline
1 & 3XMM J185246.6+003317 & G033.6+00.1 & SGR & N & 4.10e+14 & 11.55871 & 1.40e-13 & 11.5342 & 4.4 - 6.7 & 1308.85\\
2 & AXP 1E 1841-045 & G027.4+00.0 & AXP & Y & 7.00e+14 & 11.78898 & 4.09e-11 & 9.7795 & 0.75 - 2.1 & 4.57\\
3 & AXP 1E 2259+586 & G109.1-01.0 & AXP & Y & 5.90e+13 & 6.97904 & 4.84e-13 & 6.8028 & 8.8 - 14.0 & 228.59\\
4 & AXS J1617-5055  & G332.4-00.4 & PSR & Y & 3.10e+12 & 0.06936 & 1.35e-13 & 0.054 & 2.0 - 4.4 & 8.14\\
5 & CXOU J171405.7-381031 & G348.7+00.3 & AXP & Y & 5.00e+14 & 3.82535 & 6.40e-11 & 0.002 & 0.65 - 16.8 & 0.95\\
\\
6 & J0002+6216 & G116.9+00.2 & PSR & Y & 8.40e+11 & 0.11536 & 5.97e-15 & 0.1129 & 7.5 - 18.1 & 306.34\\
7 & J0007+7303 & G119.5+10.2 & PSR & Y & 1.08e+13 & 0.31587 & 3.60e-13 & 0.0808 & 13.0 - 13.0 & 13.91\\
8 & J0205+6449 & G130.7+03.1 & PSR & Y & 3.61e+12 & 0.06572 & 1.94e-13 & 0.0388 & 0.0 - 7.0 & 5.37\\
9 & J0215+6218 & G132.7+01.3 & PSR & Y & 6.10e+11 & 0.54888 & 6.62e-16 & 0.5483 & 25.0 - 33.0 & 13144.0\\
10 & J0502+4654  & G160.9+02.6 & PSR & N & 1.91e+12 & 0.63857 & 5.58e-15 & 0.6375 & 2.6 - 9.2 & 1814.18\\
\\
11 & J0534+2200 & CRAB & PSR & Y & 3.79e+12 & 0.03339 & 4.21e-13 & 0.0163 & 0.96 - 0.96 & 1.26\\
12 & J0538+2817 & G180.0-01.7 & PSR & Y & 7.33e+11 & 0.14316 & 3.67e-15 & 0.1396 & 26.0 - 34.0 & 618.38\\
13 & J0630-2834 & G276.5+19.0 & PSR & Y & 3.01e+12 & 1.24442 & 7.12e-15 & 0.002 & 1000.0 - 6000.0 & 2770.73\\
14 & J0821-4300 & G260.4-03.4 & CCO & Y & 3.27e+10 & 0.1128 & 9.28e-18 & 0.1128 & 2.2 - 5.4 & 192693.62\\
15 & J0835-4510  & G263.9-03.3 & PSR & Y & 3.38e+12 & 0.08933 & 1.25e-13 & 0.002 & 9.0 - 27.0 & 11.33\\
\\
16 & J0855-4644 & G266.2-01.2 & PSR & Y & 6.94e+11 & 0.06469 & 7.26e-15 & 0.0638 & 2.4 - 5.1 & 141.25\\
17 & J0953+0755 & G276.5+19.0 & PSR & N & 2.44e+11 & 0.25307 & 2.30e-16 & 0.2263 & 1000.0 - 6000.0 & 17442.67\\
18 & J1016-5857 & G284.3-01.8 & PSR & Y & 2.98e+12 & 0.10739 & 8.08e-14 & 0.0778 & 10.0 - 10.0 & 21.07\\
19 & J1101-6101 & G290.1-00.8 & PSR & N & 7.42e+11 & 0.0628 & 8.56e-15 & 0.0586 & 10.0 - 20.0 & 116.3\\
20 & J1105-6107 & G290.1-00.8 & PSR & Y & 1.01e+12 & 0.0632 & 1.58e-14 & 0.0552 & 10.0 - 20.0 & 63.41\\
\\
21 & J1119-6127 & G292.2-00.5 & PSR & Y & 4.10e+13 & 0.40796 & 4.02e-12 & 0.002 & 4.2 - 7.1 & 1.61\\
22 & J1124-5916 & G292.0+01.8 & PSR & Y & 1.02e+13 & 0.13548 & 7.53e-13 & 0.002 & 2.93 - 3.05 & 2.85\\
23 & J1157-6224  & G296.8-00.3 & PSR & Y & 1.27e+12 & 0.40053 & 3.93e-15 & 0.3997 & 2.0 - 11.0 & 1615.65\\
24 & J1210-5226 & G296.5+10.0 & CCO & Y & 9.83e+10 & 0.42413 & 2.22e-17 & 0.4241 & 7.0 - 10.0 & 302869.19\\
25 & J1322-6329 & G306.3-00.9 & PSR & Y & 5.60e+12 & 2.76421 & 1.11e-14 & 2.7611 & 2.5 - 15.3 & 3947.81\\
\\
26 & J1400-6325 & G310.6-01.6 & PSR & Y & 1.11e+12 & 0.03118 & 3.89e-14 & 0.0295 & 0.7 - 2.0 & 12.71\\
27 & J1513-5908 & G320.4-01.2 & PSR & Y & 1.54e+13 & 0.15158 & 1.53e-12 & 0.002 & 1.9 - 1.9 & 1.57\\
28 & J1614-5048 & G332.4+00.1 & PSR & Y & 1.08e+13 & 0.23169 & 4.95e-13 & 0.1083 & 3.0 - 8.6 & 7.42\\
29 & J1622-4944  & G333.9+00.0 & PSR & N & 4.33e+12 & 1.07297 & 1.71e-14 & 1.0713 & 0.0 - 6.0 & 994.72\\
30 & J1622-4950  & G333.9+00.0 & HBRP & Y & 1.11e+14 & 4.32702 & 2.78e-12 & 4.0555 & 0.0 - 6.0 & 24.67\\
\\
31 & J1632-4757 & G336.4+00.2 & PSR & Y & 1.88e+12 & 0.22857 & 1.51e-14 & 0.2238 & 10.0 - 10.0 & 239.97\\
32 & J1640-4631 & G338.3-00.0 & PSR & Y & 1.44e+13 & 0.20644 & 9.76e-13 & 0.002 & 1.0 - 8.0 & 3.35\\
33 & J1640-4631 & G338.5+00.1 & PSR & N & 1.44e+13 & 0.20644 & 9.76e-13 & 0.002 & 1.1 - 17.0 & 3.35\\
34 & J1702-4128 & G344.7-00.1 & PSR & Y & 3.12e+12 & 0.18214 & 5.23e-14 & 0.1746 & 3.0 - 6.0 & 55.21\\
35 & J1721-3532  & G351.7+00.8 & PSR & Y & 2.69e+12 & 0.28042 & 2.52e-14 & 0.252 & 0.0 - 68.0 & 176.41\\
\\
36 & J1747-2809 & G000.9+00.1 & PSR & Y & 2.88e+12 & 0.05215 & 1.56e-13 & 0.0418 & 1.9 - 1.9 & 5.3\\
37 & J1747-2958 & G359.1-00.5 & PSR & Y & 2.49e+12 & 0.09881 & 6.13e-14 & 0.0506 & 17.0 - 20.7 & 25.55\\
38 & J1801-2304 & G006.4-00.1 & PSR & Y & 6.93e+12 & 0.41583 & 1.13e-13 & 0.2658 & 33.0 - 36.0 & 58.34\\
39 & J1803-2137 & G008.7-00.1 & PSR & Y & 4.29e+12 & 0.13367 & 1.34e-13 & 0.002 & 15.0 - 28.0 & 15.81\\
40 & J1809-1917 & G011.2-00.3 & PSR & N & 1.47e+12 & 0.08276 & 2.55e-14 & 0.0812 & 1.4 - 2.4 & 51.45\\
\\
41 & J1809-1943 & G011.2-00.3 & HBRP & N & 1.27e+14 & 5.54074 & 2.83e-12 & 5.3685 & 1.4 - 2.4 & 31.04\\
42 & J1809-2332 & G007.5-01.7 * & PSR & Y & 2.27e+12 & 0.14679 & 3.44e-14 & 0.075 & 50.0 - 50.0 & 67.65\\
43 & J1811-1925 & G011.2-00.3 & PSR & Y & 1.71e+12 & 0.06467 & 4.40e-14 & 0.062 & 1.4 - 2.4 & 23.3\\
44 & J1813-1749 & G012.8-00.0 & PSR & Y & 2.41e+12 & 0.04474 & 1.27e-13 & 0.0396 & 1.2 - 1.2 & 5.58\\
45 & J1833-0827 & G023.3-00.3 & PSR & Y & 8.95e+11 & 0.08529 & 9.18e-15 & 0.0292 & 60.0 - 200.0 & 147.28\\
\\
46 & J1833-1034 & G021.5-00.9 & PSR & Y & 3.58e+12 & 0.06188 & 2.02e-13 & 0.0501 & 1.55 - 1.8 & 4.86\\
47 & J1846-0258 & G029.7-00.3 & HBRP & Y & 4.88e+13 & 0.32657 & 7.11e-12 & 0.002 & 1.69 - 1.85 & 0.73\\
48 & J1852+0040 & G033.6+00.1 & CCO & Y & 3.05e+10 & 0.10491 & 8.68e-18 & 0.1049 & 4.4 - 6.7 & 191609.17\\
49 & J1853-0004 & G032.8-00.1 & PSR & Y & 7.61e+11 & 0.10144 & 5.57e-15 & 0.099 & 5.7 - 22.0 & 288.7\\
50 & J1856+0113 & G034.7-00.4 & PSR & Y & 7.55e+12 & 0.26744 & 2.08e-13 & 0.2051 & 7.9 - 8.9 & 20.38\\
\\
\hline
\end{tabular}
\end{table*}

\begin{table*}
\centering
\caption{Catalogue of NSs used in our analysis.}
\label{t:catalogue1}
\begin{tabular}{cllcccrrcccrrc}
\hline
\# & NS & SNR & Type & Included & $B$ & $P$ & $\dot P$ & $P_0$ & $t_\mathrm{SNR}$ range & $\tau$  \\
   &    &     &       & Y/N      & (G) & (s) & (s/s)    & (s)   &  (Kyr)                   & (Kyr)   \\
\hline
51 & J1857+0143 & G035.6-00.4 & PSR & Y & 2.11e+12 & 0.13976 & 3.12e-14 & 0.1375 & 2.3 - 2.3 & 71.01\\
52 & J1857+0210 & G035.6-00.4 & PSR & N & 3.01e+12 & 0.63098 & 1.40e-14 & 0.63 & 2.3 - 2.3 & 714.49\\
53 & J1857+0212 & G035.6-00.4 & PSR & N & 4.14e+12 & 0.41582 & 4.03e-14 & 0.4129 & 2.3 - 2.3 & 163.57\\
54 & J1906+0722 & G041.1-00.3 & PSR & Y & 2.02e+12 & 0.11152 & 3.59e-14 & 0.1077 & 1.35 - 5.3 & 49.25\\
55 & J1913+1011 & G044.5-00.2 * & PSR & Y & 3.52e+11 & 0.03591 & 3.37e-15 & 0.0161 & 70.0 - 200.0 & 168.92\\
\\
56 & J1930+1852 & G054.1+00.3 & PSR & N & 1.03e+13 & 0.13686 & 7.51e-13 & 0.078 & 1.5 - 2.4 & 2.89\\
57 & J1930+1852 & G053.4+00.0 & PSR & Y & 1.03e+13 & 0.13686 & 7.51e-13 & 0.002 & 2.0 - 5.0 & 2.89\\
58 & J1932+1916  & G054.4-00.3 & PSR & Y & 4.46e+12 & 0.20821 & 9.32e-14 & 0.002 & 61.0 - 61.0 & 35.42\\
59 & J1952+3252 & G069.0+02.7 * & PSR & Y & 4.86e+11 & 0.03953 & 5.84e-15 & 0.0262 & 60.0 - 60.0 & 107.31\\
60 & J1957+2831 & G065.1+00.6 & PSR & Y & 9.90e+11 & 0.30768 & 3.11e-15 & 0.2987 & 40.0 - 140.0 & 1568.38\\
\\
61 & J2021+4026 & G078.2+02.1 & PSR & Y & 3.85e+12 & 0.26532 & 5.47e-14 & 0.2437 & 8.0 - 16.0 & 76.89\\
62 & J2047+5029 & G089.0+04.7 & PSR & Y & 1.38e+12 & 0.44594 & 4.18e-15 & 0.4444 & 4.8 - 18.0 & 1691.27\\
63 & J2229+6114 & G106.3+02.7 & PSR & Y & 2.03e+12 & 0.05162 & 7.83e-14 & 0.0253 & 3.9 - 12.0 & 10.45\\
64 & J2337+6151 & G114.3+00.3 & PSR & Y & 9.91e+12 & 0.49537 & 1.93e-13 & 0.446 & 7.7 - 7.7 & 40.69\\
65 & SGR 0501+4516  & G160.9+02.6 & SGR & Y & 1.90e+14 & 5.76207 & 5.94e-12 & 4.5236 & 2.6 - 9.2 & 15.38\\
\\
66 & SGR 1627-41 & G337.0-00.1 & SGR & Y & 2.20e+14 & 2.59458 & 1.90e-11 & 0.002 & 5.0 - 5.0 & 2.16\\
67 & SGR 1935+2154 & G057.2+00.8 & SGR & Y & 2.20e+14 & 3.25 & 1.43e-11 & 0.002 & 16.0 - 95.0 & 3.6\\
68 & Swift J1834.9-0846 & G023.3-00.3 & SGR & N & 1.40e+14 & 2.4823 & 7.96e-12 & 0.002 & 60.0 - 200.0 & 4.94\\
\hline
\end{tabular}
\end{table*}

\section{Testing the maximum likelihood technique}
\label{s:synthetic}

Here we construct simulations to test the maximum likelihood approach. We draw 60 objects which is comparable to size of our catalogue. The initial spin periods and magnetic fields are drawn from log-normal distributions with the following parameters $\mu_B = 12.44$, $\sigma_B = 0.44$ and $\mu_p = -1.04$ and $\sigma_p = 0.42$. Actual ages of radio pulsars $t$ are drawn from uniform distribution in range $10^3$~--~$10^5$~years. These actual ages are not equal to the measured ages $t'$ which we assumed to be drawn from another uniform distribution centred at actual ages $t' \sim U (0.5 t, 1.5 t)$. We compute new periods as:
\begin{equation}
P = \sqrt{P_0^2 + 2 \kappa B_0^2 t}    
\end{equation}
where $\kappa = 9.77\times 10^{-40}$~s~G$^{-2}$ and it is simply inverse of squared numerical coefficient from eq. (\ref{eq:bp}). 

\begin{figure*}
    \begin{minipage}{0.49\linewidth}
    \includegraphics[width=\columnwidth]{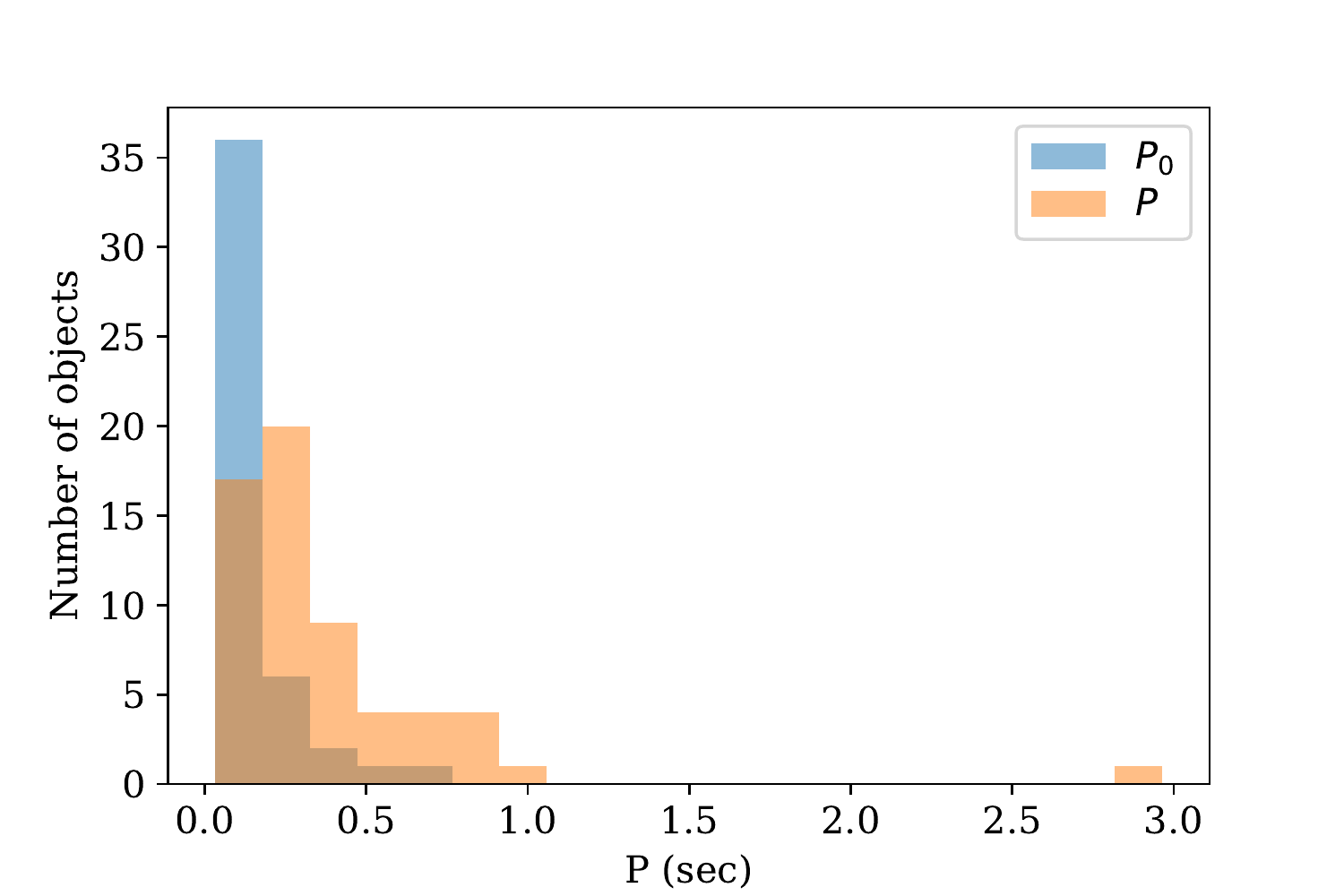}
    \end{minipage}
	\begin{minipage}{0.49\linewidth}
    \includegraphics[width=\columnwidth]{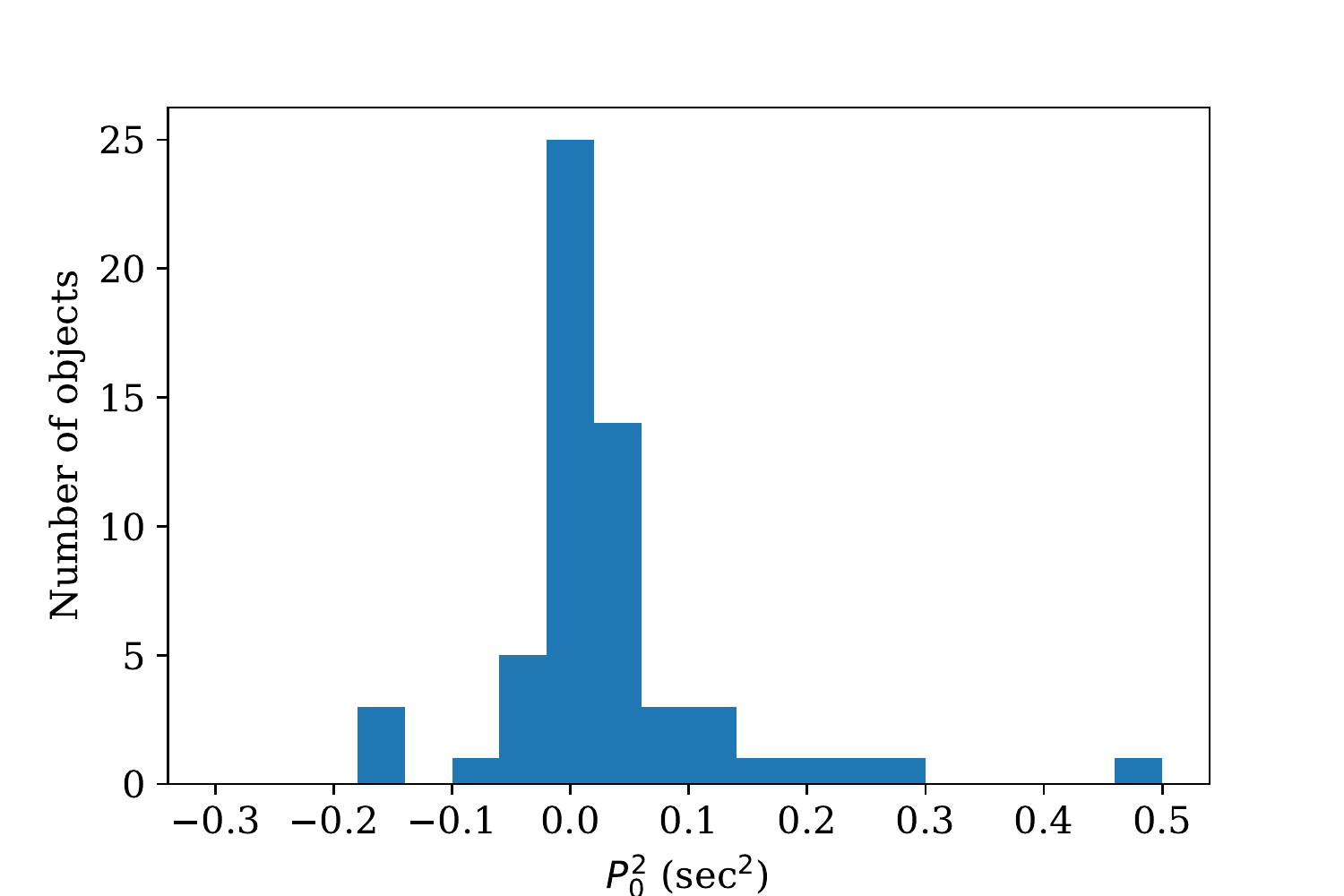}
    \end{minipage}
    \caption{Left panel: the histograms for simulated initial periods and periods computed at some ages ranging $10^3$~--~$10^5$~years. Right panel: estimated $P_0^2$ using measured ages of synthetic radio pulsars.}
    \label{fig:P_simul}
\end{figure*}

We show the distribution of initial periods and computed instantaneous periods result in Figure~\ref{fig:P_simul}, left panel. If we formally compute initial periods using measured ages:
\begin{equation}
P_0^2 = P^2 - 2 \kappa B_0^2 t'  
\end{equation}
in some cases we obtain negative $P_0^2$ because $t' > t$ i.e.
ages of some SNRs are overestimated in comparison to their actual ages. We show the distribution of computed $P_0^2$ in right panel of Figure~\ref{fig:P_simul}. Therefore, already uncertainty in measured SNR ages could produce negative period which we see in our data. 

\begin{figure}
    \includegraphics[width=\columnwidth]{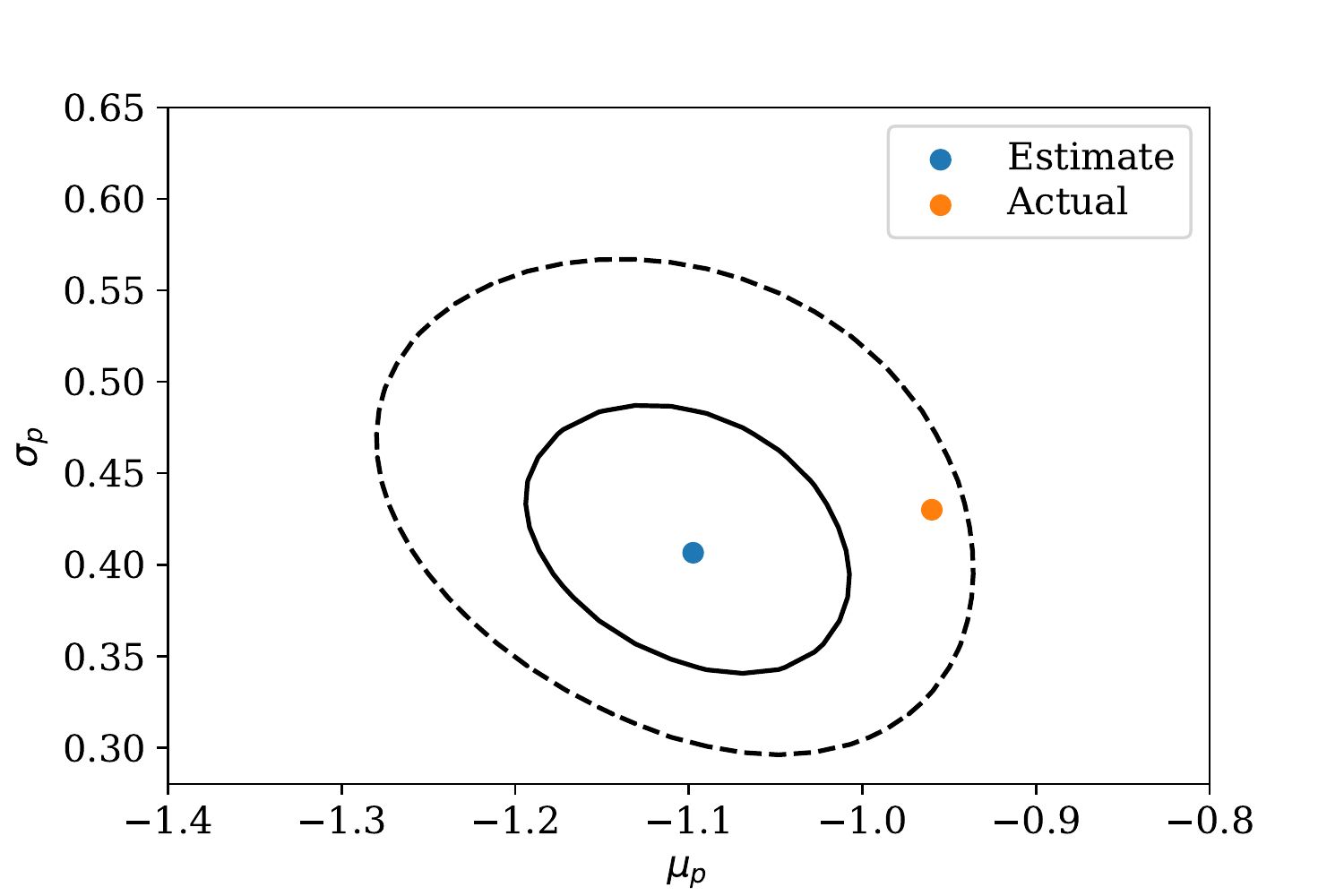}
    \caption{Contours of constant likelihood for the initial periods drawn from the log-normal distribution. Blue dot corresponds to the maximum likelihood, solid and dashed contours to 68 and 99 percent confidence intervals. Orange dot corresponds to parameters of the assumed distribution.}
    \label{fig:P_simul_rest}
\end{figure}

We apply our maximum likelihood technique to synthetic data and check that we successfully restore the parameters of the initial distribution, see Figure~\ref{fig:P_simul_rest}. Confidence interval (99\%) estimated for the restored value $\mu_p = -1.1$  $\sigma_p = 0.41$ includes actual value. If we increase the number of objects in our synthetic catalogue, the confidence interval shrinks as it is expected. Thus our maximum likelihood technique could successfully estimate the parameters of initial period distribution even when some ages are overestimated. 



\bsp	
\label{lastpage}
\end{document}